\documentclass[11pt,a4paper]{article}
\usepackage{jheppub}
\usepackage{amsmath,amssymb,bm}
\usepackage{graphicx}
\usepackage{hepunits}
\usepackage{braket}
\usepackage{slashed}
\usepackage{diagbox}
\usepackage[svgnames]{xcolor}
\usepackage{multirow}

\usepackage{subfigure}

\allowdisplaybreaks

\ifcsname SI \endcsname
  \renewcommand{\unit}{\SI}
\fi

\DeclareUnicodeCharacter{2212}{-}

\newcommand{\diff}{\mathrm{d}}

\title{Thrust distribution in Higgs decays up to the fifth logarithmic order}

\author[a]{Wan-Li Ju,}
\emailAdd{Wanli.Ju@mi.infi.it}
\author[b]{Yongqi Xu,}
\emailAdd{xuyongqi@pku.edu.cn}
\author[c]{Li Lin Yang,}
\emailAdd{yanglilin@zju.edu.cn}
\author[d]{Bin Zhou}
\emailAdd{zb0429@sjtu.edu.cn}

\affiliation[a]{INFN, Sezione di Milano, Via Celoria 16, 20133 Milano, Italy}
\affiliation[b]{School of Physics and State Key Laboratory of Nuclear Physics and Technology, Peking University, Beijing 100871, China}
\affiliation[c]{Zhejiang Institute of Modern Physics, School of Physics, Zhejiang University, Hangzhou 310027, China}
\affiliation[d]{INPAC, Shanghai Key Laboratory for Particle Physics and Cosmology,
School of Physics and Astronomy, Shanghai Jiao Tong University, Shanghai 200240, China}

\abstract{
In this work, we extend the resummation for the thrust distribution in Higgs decays up to the fifth logarithmic order. We show that one needs the accurate values of the three-loop soft functions for reliable predictions in the back-to-back region. This is especially true in the gluon channel, where the soft function exhibits poor perturbative convergence.
}

\begin{document}

\maketitle

\clearpage

\section{Introduction}\label{sec:introduction}

The hadronic decays of the Higgs boson provide a unique windows to study the Yukawa couplings of the lighter quarks such as the charm quark and the strange quark. These rare decays might be enhanced by new physics effects beyond the standard model (see, e.g., \cite{Giudice:2008uua, Bishara:2015cha}), and can be probed at the Large Hadron Collider (LHC) and the future Higgs factories \cite{ILC:2013jhg, Moortgat-Pick:2015lbx, CEPCStudyGroup:2018ghi, Bambade:2019fyw, FCC:2018evy}. However, the hadronic decays of the Higgs boson can also proceed through the $H \to gg$ partonic channel. While the gluon channel is also useful, it is desirable to distinguish it from the $H \to q\bar{q}$ channel to gain maximal information about the Yukawa couplings. To this end, it is important to study various differential distributions in these two channels.

A classical differential distribution for hadronic final states is the event shape variable ``thrust'' \cite{Farhi:1977sg}. It was extensively studied in the process $e^+e^- \to \text{hadrons}$. In the context of $H \to \text{hadrons}$, the next-to-leading order (NLO) and approximate next-to-next-to-leading order (NNLO) predictions were calculated in \cite{Gao:2019mlt}. These fixed-order results suffer from large logarithms in the endpoint region, which need to be resummed to all orders in the strong coupling $\alpha_s$. The resummation framework is also well-established for $e^+e^- \to \text{hadrons}$ \cite{Catani:1992ua, Schwartz:2007ib, Becher:2008cf, Dissertori:2009ik, Abbate:2010xh, Monni:2011gb, Banfi:2014sua, Hoang:2014wka, Bell:2018gce}. The applications to the Higgs case were carried out in \cite{Mo:2017gzp, Alioli:2020fzf} at the next-to-next-to-leading logarithmic (NNLL and NNLL$'$) accuracies. In this work, we extend the resummation accuracy up to the fifth order, and present the results at N$^3$LL$'$ and N$^4$LL.

The paper is organized as follows. In Section~\ref{sec:Theoretical framework} we briefly review the factorization formula for the thrust distribution, and give technical details of the resummation framework.
In Section~\ref{sec:Numeric results} we provide numeric results for the resummed thrust distributions, with jet and soft scales chosen in the Laplace space. The summary and outlook come in Section~\ref{sec:summary}. The alternative results with jet and soft scales chosen in the momentum space are presented in Appendix~\ref{sec:app:momentum}, and we leave some lengthy expressions to the remaining Appendices.

\section{Theoretical framework}
\label{sec:Theoretical framework}

We consider the process $H \to \text{hadrons}$ induced by the following effective Lagrangian
\begin{align}
\mathcal{L}_{\text{eff}} &= \frac{\alpha_s(\mu) C_t(m_t,\mu)}{12 \pi v} O_g + \sum_q \frac{y_q(\mu)}{\sqrt{2}} O_q \nonumber
\\
&\equiv \frac{\alpha_s(\mu) C_t(m_t,\mu)}{12 \pi v} H G^{\mu\nu,a} G_{\mu\nu}^{a} + \sum_q \frac{y_q(\mu)}{\sqrt{2}} H \bar{\psi}_q \psi_q \, ,
\label{eq:Leff}
\end{align}
where $v$ is the Higgs vacuum expectation value; $H$ represents the physical Higgs boson after electroweak symmetry breaking; $G_{\mu\nu}^a$ is the field strength tensor of the gluon field; $\psi_q$ represent the light quark fields. We will ignore the masses of the light quarks, but keep the Yukawa couplings $y_q$ non-vanishing. The strong coupling $\alpha_s$, the Yukawa coupling $y_q$ and the Wilson coefficient $C_t$ of the effective operator are renormalized in the $\overline{\text{MS}}$ scheme at the scale $\mu$.

The thrust variable $T$ is defined as
\begin{equation}
T \equiv \max_{\vec{n}} \frac{\sum_i \vert \vec{n} \cdot \vec{p}_i \vert}{\sum_i \vert \vec{p}_i \vert} \, ,
\end{equation}
where $\vec{p}_i$ denote the 3-momenta of final state particles. The unit vector $\vec{n}$ that maximize the above ratio is called the thrust axis. For convenience we introduce the variable $\tau \equiv 1 - T$. In this work we are concerned with the limit $T \to 1$ or $\tau \to 0$. Physically this corresponds to two back-to-back jets in the final state. In this limit the differential decay rate can be factorized into the product (convolution) of a hard function, a soft function and two jet functions \cite{Catani:1991kz,Catani:1992ua,Contopanagos:1996nh,Kidonakis:1998bk,Berger:2003iw,Schwartz:2007ib,Becher:2008cf,Gao:2019mlt}:
\begin{align}
\frac{\diff\Gamma^q}{\diff\tau} &= \Gamma_{B}^{q}(\mu) \, \vert C_S^q (m_H,\mu) \vert^2 \int \diff p^2_{n} \, \diff p^2_{\bar{n}} \, \diff k \, \delta \left( \tau - \frac{p^2_{n}+p^2_{\bar{n}}}{m_H^2} - \frac{k}{m_H} \right) \nonumber
\\
&\hspace{7cm} \times J_{n}^q(p^2_{n},\mu) \, J_{\bar{n}}^q(p^2_{\bar{n}},\mu) \, S^q(k,\mu) \, , \nonumber
\\
\frac{\diff\Gamma^g}{\diff\tau} &= \Gamma_{B}^{g}(\mu) \, \vert C_t (m_t,\mu) \vert^2 \, \vert C_S^g (m_H,\mu) \vert^2 \int \diff p^2_{n} \, \diff p^2_{\bar{n}} \, \diff k \, \delta \left( \tau - \frac{p^2_{n}+p^2_{\bar{n}}}{m_H^2} - \frac{k}{m_H} \right) \nonumber
\\
&\hspace{7cm} \times J_{n}^g(p^2_{n},\mu) \, J_{\bar{n}}^g(p^2_{\bar{n}},\mu) \, S^g(k,\mu) \, ,
\label{eq: factorization formula}
\end{align}
where the superscript $q$ or $g$ labels the partonic subprocesses $H \to q\bar{q}$ or $H \to gg$, with $\Gamma_{B}^{q}$ and $\Gamma_{B}^{g}$ being the corresponding total decay rates at the Born level. Their explicit expressions are
\begin{equation}
\Gamma_{B}^{q} = \frac{y_q^2(\mu) \, m_H \,  C_A}{16 \pi} \, , \quad \Gamma_{B}^{g} = \frac{\alpha_s^2(\mu) \, m_H^3}{ 72 \, \pi^3 \, v^2} \, .
\label{eq: Born cross section}
\end{equation}
In the factorization formula, $C_S^i$ (with $i=q,g$) are hard Wilson coefficients arising when matching the full theory of QCD to the soft-collinear effective theory (SCET) \cite{Bauer:2000ew, Bauer:2000yr, Bauer:2001ct, Bauer:2002nz, Bauer:2001yt, Beneke:2002ph}; $S^i$ are soft functions defined as the vacuum expectation values of soft Wilson-loop operators; $J_n^i$ and $J_{\bar{n}}^i$ are jet functions along the two light-like directions $n^\mu = (1,\vec{n})$ and $\bar{n}^\mu = (1,-\vec{n})$, where $\vec{n}$ is the thrust axis.

The various ingredients satisfy renormalization group (RG) equations
\begin{align}
\label{eq:evolution equations in momentum space}
\frac{\diff}{\diff\ln\mu} y_q(\mu) &= \gamma_y(\alpha_s(\mu)) \, y_q(\mu) \, , \nonumber
\\
\frac{\diff}{\diff\ln\mu} C_t(m_t,\mu) &= \gamma_t (\alpha_s(\mu)) \, C_t(\mu^2) \, , \nonumber
\\
\frac{\diff}{\diff\ln\mu} C_S^{i}(m_H,\mu) &= \left[ \Gamma^{i}_{\text{cusp}}(\alpha_s(\mu)) \ln \frac{-m_H^2-i \epsilon}{\mu^2} + \gamma^{i}_H(\alpha_s(\mu)) \right]  C_S^{i}(m_H,\mu) \, , \nonumber
\\
\frac{\diff}{\diff\ln\mu} J^i(p^2,\mu) &= \left[ -2 \Gamma^{i}_{\text{cusp}}(\alpha_s(\mu)) \ln\frac{p^2}{\mu^2} - 2\gamma^{i}_J(\alpha_s(\mu)) \right] J^{i}(p^2,\mu) \nonumber
\\
&\hspace{2cm} +2 \Gamma^{i}_{\text{cusp}}(\alpha_s(\mu)) \int^{p^2}_0 \diff q^2 \,  \frac{J^i(p^2,\mu)-J^i(q^2,\mu)}{p^2-q^2}\, , \nonumber
\\
\frac{\diff}{\diff\ln\mu} S^{i}(k,\mu) &=\left[ 4\Gamma^{i}_{\text{cusp}}(\alpha_s(\mu)) \ln\frac{k}{\mu} - 2\gamma^{i}_S(\alpha_s(\mu)) \right] S^{i}(k,\mu) \nonumber
\\
&\hspace{2cm} -4 \Gamma^{i}_{\text{cusp}}(\alpha_s(\mu)) \int^{k}_0 \diff q \, \frac{S^i(k,\mu)-S^i(q,\mu)}{k-q} \, .
\end{align}
Note that the evolution equations for the jet and soft functions involve convolutions. It is useful to introduce the Laplace transformed functions
\begin{align}
\tilde{j}^i(L_J,\mu) &= \int_0^\infty \diff p^2 \, \exp \left( -\frac{N p^2}{m_H^2} \right) J^i(p^2,\mu) \, , \nonumber
\\
\tilde{s}^i(L_S,\mu) &= \int_0^\infty \diff k \, \exp \left( -\frac{N k}{m_H} \right) S^i(k,\mu) \, ,
\end{align}
where
\begin{equation}
L_J = \ln\frac{m_H^2}{\mu^2 \bar{N}} \, , \quad L_S = \ln\frac{m_H}{\mu \bar{N}} \, ,
\end{equation}
with $\bar{N} \equiv N e^{\gamma_E}$ and $\gamma_E$ being the Euler's constant. The Laplace-space jet and soft functions satisfy local RG equations
\begin{align}
\frac{\diff}{\diff\ln\mu} \tilde{j}^i(L_J,\mu) &= \left[ -2 \Gamma^{i}_{\text{cusp}}(\alpha_s(\mu)) L_J - 2\gamma^{i}_J(\alpha_s(\mu)) \right] \tilde{j}^{i}(L_J,\mu) \, , \nonumber
\\
\frac{\diff}{\diff\ln\mu} \tilde{s}^{i}(L_S,\mu) &= \left[ 4 \Gamma^{i}_{\text{cusp}}(\alpha_s(\mu)) L_S - 2\gamma^{i}_S(\alpha_s(\mu)) \right] \tilde{s}^{i}(L_S,\mu) \, .
\end{align}

Under the Laplace transform, the differential decay rates are expressed as
\begin{align}
\int^{\infty}_{0}\diff\tau \, e^{-\tau N} \, \frac{\diff{\Gamma}^{q}}{\diff\tau} &= \Gamma_B^{q}(\mu) \, \vert C_S^q (m_H, \mu) \vert^2 \left[ \tilde{j}^q(L_J,\mu) \right]^2 \tilde{s}^q(L_S,\mu) \, , \nonumber
\\
\int^{\infty}_{0}\diff\tau \, e^{-\tau N} \, \frac{\diff{\Gamma}^{g}}{\diff\tau} &=  \Gamma_B^{g}(\mu) \, \vert C_t (m_t,\mu) \vert^2 \, \vert C_S^g (m_H,\mu) \vert^2 \left[ \tilde{j}^g(L_J,\mu) \right]^2 \tilde{s}^g(L_S,\mu) \, .
\label{eq:Laplace-transformed thrust distributions}
\end{align}
For small $\tau$, the dominant contribution arises from the region of large $N$. In this case the large logarithms $L_J$ and $L_S$ appear with increasing powers at each order in the perturbative expansions of the jet and soft functions. We will resum these logarithms to all orders in $\alpha_s$ using RG evolution.

In the RG equations, $\Gamma^{i}_{\text{cusp}}$ are the cusp anomalous dimensions, which are known up to the four-loop accuracy \cite{Moch:2004pa,Vogt:2004mw, Henn:2019swt, vonManteuffel:2020vjv}, and their fifth order contributions were estimated in \cite{Herzog:2018kwj}. The beta function governing the running strong coupling is known up to the five-loop order in \cite{Caswell:1974gg,Jones:1974mm,Tarasov:1980au,Larin:1993tp,vanRitbergen:1997va, Czakon:2004bu}. The anomalous dimension $\gamma^y$ for the Yukawa coupling is the same as that for the quark masses in the $\overline{\text{MS}}$ scheme, and is known up to the fifth order in \cite{Tarrach:1980up,Larin:1993tq,Vermaseren:1997fq, Chetyrkin:1997dh,Baikov:2014qja}. The non-cusp anomalous dimension $\gamma_t$ governing scale evolution of the Wilson coefficient $C_t$ is also known on the fifth level \cite{Spiridonov:1984br, Kramer:1996iq, Chetyrkin:1997un, Schroder:2005hy, Chetyrkin:2005ia,Liu:2015fxa}. The up-to four-loop results for the remaining anomalous dimensions can be found for $\gamma^i_H$ in \cite{Dawson:1990zj,Djouadi:1991tka,Harlander:2000mg,Moch:2005tm,Gehrmann:2005pd,Baikov:2009bg,Gehrmann:2010ue,Gehrmann:2014vha,Chakraborty:2022yan, Lee:2022nhh}, for $\gamma^i_S$ in \cite{Korchemsky:1993uz,Belitsky:1998tc,Li:2014afw, Duhr:2022cob}, and for $\gamma^i_J$ in \cite{Bauer:2003pi,Bosch:2004th,Becher:2006qw,Becher:2009th,Bruser:2018rad,Banerjee:2018ozf,Becher:2010pd, Duhr:2022cob}. The Wilson coefficient $C_t$ is known up to the four-loop order \cite{Inami:1982xt, Djouadi:1991tk, Chetyrkin:1997iv, Chetyrkin:1997un, Chetyrkin:2005ia, Schroder:2005hy, Baikov:2016tgj} and so is the hard sector \cite{Harlander:2000mg,Gehrmann:2005pd,Baikov:2009bg,Ahmed:2014pka,Gehrmann:2010ue, Gehrmann:2014vha,Lee:2022nhh,Chakraborty:2022yan} in the factorization of Eq.~\eqref{eq: factorization formula}. As for the fixed-order expansions of   the jet functions and the soft functions, the analytic results are known up to the three-loop order \cite{Bauer:2003pi, Bosch:2004th,Becher:2006qw, Fleming:2007xt,Schwartz:2007ib,Becher:2008cf, Becher:2009th, Becher:2010pd, Bruser:2018rad, Banerjee:2018ozf, Kelley:2011ng}, with the exception of the scale-independent terms of the three-loop soft function. We collect all these ingredients in the Appendices~\ref{sec:app:fixed order ingredients} and \ref{sec:app:Anomalous dimensions}. They allow us to perform the resummation of large logarithms to the N$^3$LL$'$ order and approximately to the N$^4$LL order. For the counting of logarithmic orders, we refer to Table~\ref{tab:power counting rules}.

\begin{table}[t!]
\begin{center}
\begin{tabular}{|c|c|c|c|c|}
\hline   Logarithmic accuracy & $\Gamma_{\text{cusp}},\beta$ & $ \gamma_{t,y,H,j,s}$ & $C_t,C_S,\tilde{j},\tilde{s}$ \\
\hline   NNLL$'$ & 3-loop & 2-loop & 2-loop \\
\hline   N$^3$LL & 4-loop & 3-loop & 2-loop \\
\hline   N$^3$LL$'$ & 4-loop & 3-loop & 3-loop \\
\hline   N$^4$LL & 5-loop & 4-loop & 3-loop\\
\hline
\end{tabular}
\end{center}
\caption{Definitions of the logarithmic orders.\label{tab:power counting rules}}
\end{table}

To resum the large logarithms, we choose appropriate scales for each of the functions in the factorization formula, and use the RG equations to evolve them to a common scale. The choice of scales can be done either in the Laplace $N$-space or in the momentum $\tau$-space. In the following, we present results with scale choices in the Laplace space, while those in the momentum space will be discussed in Appendix~\ref{sec:app:momentum}. We choose the scales for the $C_t$, $C_S$, $\tilde{j}$ and $\tilde{s}$ functions to be
\begin{equation}
\mu_t = e_t m_t \, , \quad \mu_h = e_h m_H \, , \quad \mu_j = e_j \frac{m_H}{\sqrt{\bar{N}}} \, , \quad \mu_s = e_s \frac{m_H}{\bar{N}} \, ,
\label{eq:scales_Laplace}
\end{equation}
where by default we take $e_t=e_h=e_j=e_s=1$, and we vary them up and down by a factor of two to estimate the associated uncertainties. The resummed differential decay rates in the Laplace space can be written as
\begin{align}
\widetilde{\Gamma}^{q}(N) &= \Gamma_B^{q}(\mu_h) \, U^q(\mu_h,\mu_j,\mu_s) \, \vert C_S^q (m_H,\mu_h) \vert^2 \left[ \tilde{j}^q(L_J,\mu_j) \right]^2 \tilde{s}^q(L_S,\mu_s) \left( \frac{m_H}{\mu_s \bar{N}} \right)^{\eta_q} , \nonumber
\\
\widetilde{\Gamma}^{g}(N) &= \Gamma_B^{g}(\mu_h) \, U^g(\mu_t,\mu_h,\mu_j,\mu_s) \, \vert C_t(m_t,\mu_t) \vert^2 \, \vert C_S^g(m_H,\mu_h)\vert^2 \nonumber
\\
&\hspace{5cm} \times \left[ \tilde{j}^g(L_J,\mu_j) \right]^2 \tilde{s}^g(L_S,\mu_s) \left( \frac{m_H}{\mu_s \bar{N}} \right)^{\eta_g} ,
\label{eq:Laplace-transformed resummed thrust distributions}
\end{align}
where the evolution functions are given by
\begin{align}
\label{eq:evolution functions}
U^q(\mu_h,\mu_j,\mu_s) &= \exp \bigg[ 4S^q(\mu_h,\mu_j) + 4S^q(\mu_s,\mu_j) - 2A^q_{\text{cusp}}(\mu_h,\mu_j) \ln\frac{m_H^2}{\mu_h^2} \nonumber
\\
&\hspace{5cm} - 2A^q_S(\mu_h,\mu_s) - 4A^q_J(\mu_h,\mu_j) \bigg] \, , \nonumber
\\
U^g(\mu_t,\mu_h,\mu_j,\mu_s) &= \exp \bigg[ 2A_t(\mu_h,\mu_t) + 4S^g(\mu_h,\mu_j) + 4S^g(\mu_s,\mu_j) - 2A^g_{\text{cusp}}(\mu_h,\mu_j) \ln\frac{m_H^2}{\mu_h^2} \nonumber
\\
&\hspace{5cm} - 2A^g_S(\mu_h,\mu_s) - 4A^g_J(\mu_h,\mu_j) \bigg] \, ,
\end{align}
in which the functions $S^{q,g}$ and $A^{q,g}_{i}$ are defined by \cite{Becher:2006mr}
\begin{align}
S^{q,g}(\nu,\mu) &= - \int^{\alpha_{s}(\mu)}_{\alpha_{s}(\nu)} \diff\alpha_s \frac{\Gamma^{q,g}_{\text{cusp}}(\alpha_s)}{\beta(\alpha_s)}
\int^{\alpha_s}_{\alpha_s(\nu)} \frac{\diff\tilde{\alpha}_s}{\beta(\tilde{\alpha}_s)} \, , \nonumber
\\
A^{q,g}_i(\nu,\mu) &= -\int^{\alpha_{s}(\mu)}_{\alpha_{s}(\nu)}\mathrm{d}\alpha_s \dfrac{\gamma^{q,g}_i(\alpha_s)}{\beta(\alpha_s)} \, ,
\label{eq:SAdefinition}
\end{align}
for $i=\text{cusp},t,S,J$, and $\eta_{q,g}=4A^{q,g}_{\text{cusp}}(\mu_j,\mu_s)$. The momentum-space differential decay rates can then be obtained through an inverse Laplace transform
\begin{equation}
\label{eq:inverse_Laplace}
\frac{\diff {\Gamma}^{q,g}}{\diff\tau} = \frac{1}{2\pi i} \int_{-i\infty}^{+i\infty}
\diff N  \, e^{N\tau} \, \widetilde{\Gamma}^{q,g}(N) \, .
\end{equation}
The integration contour should, in principle, be chosen such that all singularities of the integrand are situated to the left side. However, the resummed integrand develops a Landau pole at large $N$ due to the scale choices $\mu_j \sim m_H/\sqrt{\bar{N}}$ and $\mu_s \sim m_H/\bar{N}$, which signals the breakdown of perturbation theory in that region. Correspondingly, the inverse Laplace transform of the perturbatively resummed integrand suffers from an ambiguity of non-perturbative origin. We adopt the so-called Minimal Prescription \cite{Catani:1996yz}, in which the contour lies to the right of all physical singularities but to the left of the Landau pole.

With the generic framework, we still need to specify a few details in the evaluation of the resummed differential decay rates at a given logarithmic accuracy. The strong coupling $\alpha_s$ at a given scale $\mu$ is evaluated according to
\begin{align}
\alpha_s(\mu) &= \frac{\alpha_s(\nu)}{X} \Bigg\{ 1 - \frac{\alpha_s(\nu)}{4\pi X} \frac{\beta_1 \ln(X)}{\beta_0}
+ \left( \frac{\alpha_s(\nu)}{4\pi X} \right)^2
\bigg[ \frac{\beta_1^2}{\beta_0^2} \left( \ln^2(X)-\ln(X)-1+X \right) \nonumber
\\
&+ \frac{\beta_2}{\beta_0} (1-X) \bigg]
+ \left( \frac{\alpha_s(\nu)}{4\pi X} \right)^3
\bigg[ \frac{\beta_1^3}{\beta_0^3} \bigg( -\frac{X^2}{2}+X-\ln^3(X)+\frac{5\ln^2(X)}{2} \nonumber
\\
&+ 2(1-X) \ln(X) - \frac{1}{2} \bigg) + \frac{\beta_3}{2\beta_0}(1-X^2) + \frac{\beta_1\beta_2}{\beta^2_0} \Big( 2 X \ln(X) - 3 \ln(X) \nonumber
\\
& - X (1 - X) \Big) \bigg]
+ \left( \frac{\alpha_s(\nu)}{4\pi X} \right)^4
\bigg[ -\frac{\beta_4 (X^3-1)}{3 \beta_0} + \frac{\beta_3 \beta_1}{6 \beta_0^2} \big( (X-1) (4X^2+X+1) \nonumber
\\
&+ 6(X^2-2) \ln(X) \big)
+\frac{\beta_2^2 (X-1)^2 (X+5)}{3 \beta_0^2}
+\frac{\beta_2 \beta_1^2}{\beta_0^3} \Big( -(X+3) (X-1)^2 \nonumber
\\
&+ \ln(X)(-2X^2+5X-3) - 3 (X-2) \ln^2(X) \big)
+\frac{\beta_1^4}{6 \beta_0^4} \Big( (X-1)^2 (2X+7) \nonumber
\\
&+ 6 \ln^4(X) - 26\ln^3(X) + 9 (2X-1) \ln^2(X) + 6 (X-4) (X-1) \ln(X) \Big) \bigg] \Bigg\} \, ,
\label{eq:alphaexp}
\end{align}
where the initial scale is chosen at the $Z$ boson mass, $\nu = m_Z$, and
\begin{equation}
X = 1 + \frac{\alpha_s(\nu)}{2\pi} \beta_0 \ln\frac{\mu}{m_Z} \, .
\end{equation}
The coefficients of the beta function are defined through
\begin{equation}
\frac{\diff\alpha_s}{\diff\ln\mu} = -2\alpha_s \sum_{n=0}^\infty \left(\frac{\alpha_s}{4\pi}\right)^{n+1} \beta_n \, .
\end{equation}
The Yukawa coupling $y_q$ at a given scale is evaluated with
\begin{equation}
y_q(\mu)=y_q(m_H) \, \exp \left[ A^{q }_y(\mu,m_H) \right] .
\label{eq:yqexp2}
\end{equation}

The evolution factors $U^{q,g}$ and the factors involving $\eta_{q,g}$ are expanded on the exponent up to a given logarithmic accuracy defined in Table~\ref{tab:power counting rules}. The expansion is done by counting the large logarithms $\ln(\nu/\mu)$ as $\mathcal{O}(1/\alpha_s)$. The fixed-order factors ($|C_{S}^{q}|^2 \left[\tilde{j}^q\right]^2 \tilde{s}^q$ and $|C_{t}|^2 |C_{S}^{g}|^2 \left[\tilde{j}^g\right]^2 \tilde{s}^g$) are also expanded up to a given loop order. We are now ready to perform numeric evaluations of the resummed differential decay rates. The results are presented in the next Section.

\section{Numeric results}
\label{sec:Numeric results}

\subsection{Choice of parameters and estimation of uncertainties}

In this section we are devoted to the numeric results. Throughout this paper, we choose $\alpha_s(m_Z)=0.1181$, $m_H=\unit{125.1}{\GeV}$ and $m_t=\unit{172.9}{\GeV}$ \cite{ParticleDataGroup:2018ovx}. The scales are chosen as in Eq.~\eqref{eq:scales_Laplace}. Note that this choice is conventional in the small-$\tau$ region considered in this work. On the other hand, if one wants to match the resummed distributions to the fixed-order ones, it is necessary to deal with the intermediate regime between the resummation dominated small-$\tau$ region and the fixed-order dominated large-$\tau$ region. We leave this subtlety to future investigations. We estimate the perturbative uncertainties by varying each of $e_t$, $e_h$, $e_j$ and $e_s$ up and down by a factor of two, while keeping the others at their defaults. The resulting variations of the differential decay rates are then added in quadrature.

The scale-independent constant terms of the three-loop soft functions are not known yet. The term in the quark channel was extracted in \cite{Bruser:2018rad} through a numeric fit to the fixed-order thrust distribution, which has a large uncertainty. In this work we set
\begin{equation}
c_{3q}^S = -19988 \pm 5000 \, ,
\end{equation}
and estimate the corresponding variation of the resummed distribution. For the gluon channel we apply the Casimir scaling and set
\begin{equation}
c_{3g}^S = -45433 \pm 11250 \, .
\end{equation}
The five-loop cusp anomalous dimensions are also unknown, with only a rough estimation available \cite{Herzog:2018kwj}. However, we have checked that they only have a rather mild effect on the resummed thrust distributions.

\subsection{The resummed thrust distributions in the gluon channel}

We now show the resummed thrust distributions in the gluon channel at various logarithmic accuracies. The baseline for comparison is the NNLL$'$ result, that is the state-of-the-art accuracy in the literature (see Refs.~\cite{Mo:2017gzp, Alioli:2020fzf}, although they adopted scale choices in the momentum space).

\begin{figure}[t!]
\begin{center}
\includegraphics[width=0.45\textwidth]{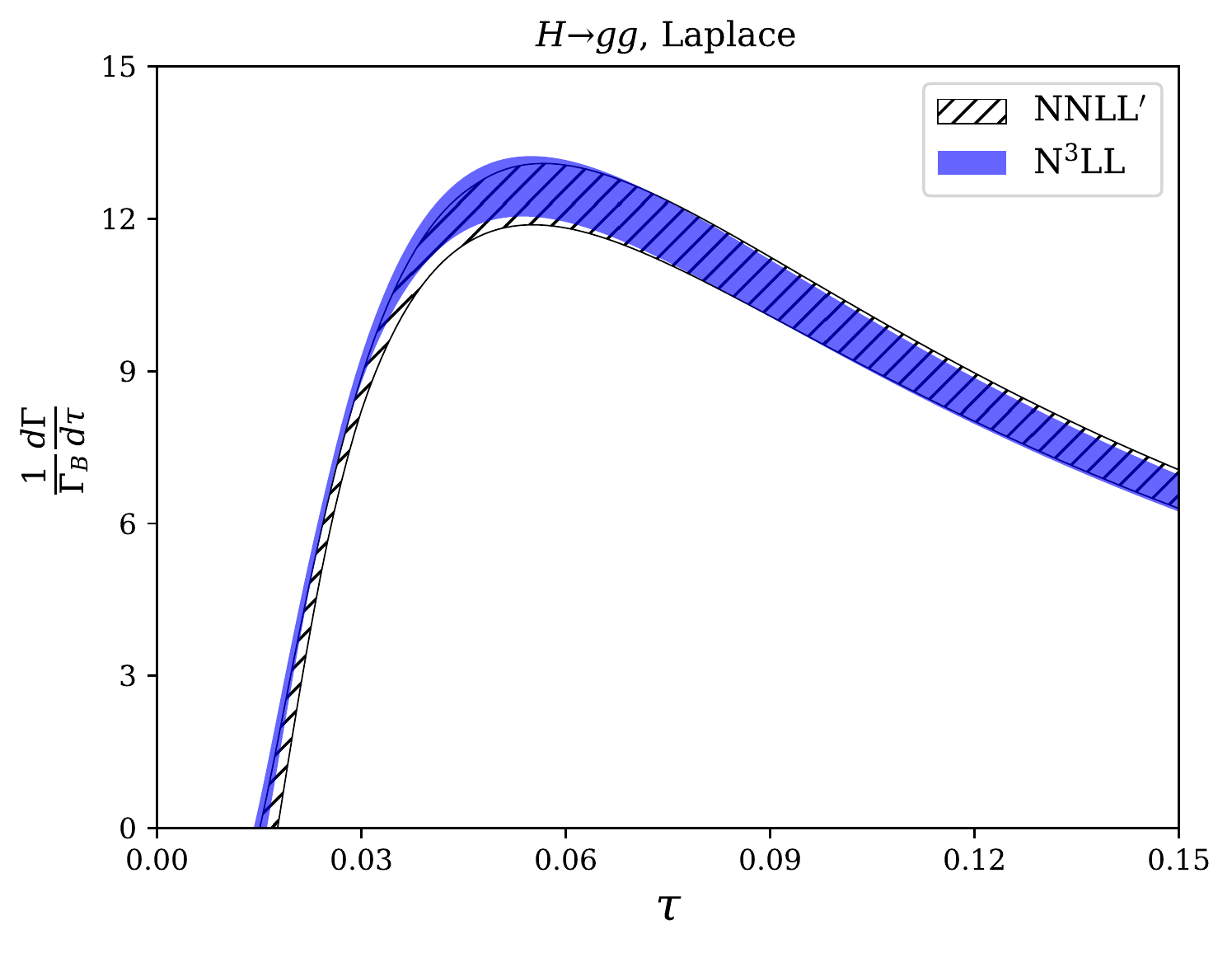}
\hspace{0.5cm}
\includegraphics[width=0.45\textwidth]{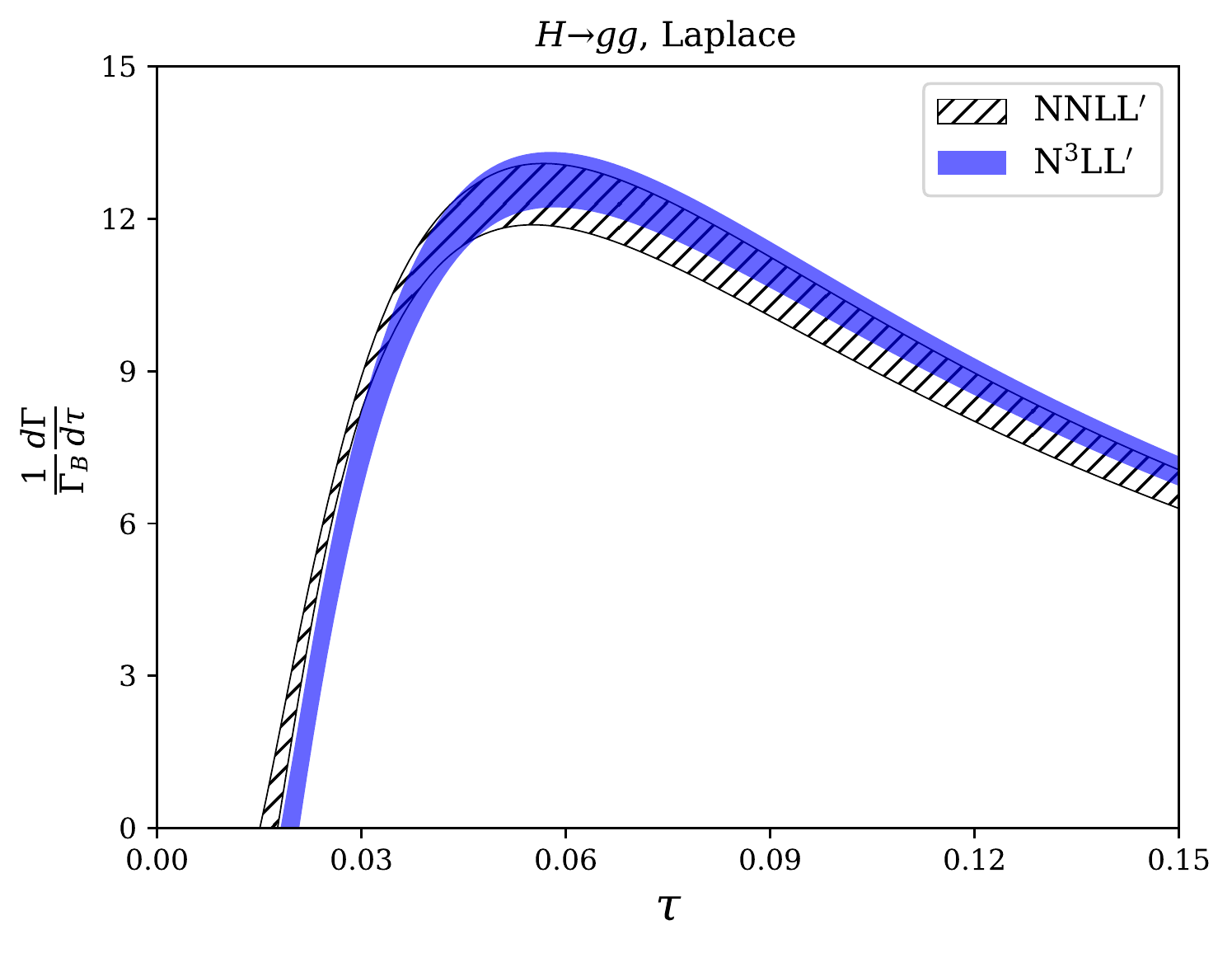}
\end{center}
\caption{\label{figure:hggn3ll}The resummed thrust distributions in the gluon channel. Left: NNLL$'$ vs. N$^3$LL; Right: NNLL$'$ vs. N$^3$LL$'$.}
\end{figure}

In Fig.~\ref{figure:hggn3ll}, we show the comparison between NNLL$'$ and N$^3$LL, and that between NNLL$'$ and N$^3$LL$'$, for the $\tau$ range $[0.01,0.15]$. This covers the small-$\tau$ and intermediate-$\tau$ regions, but cuts out the large-$\tau$ region where fixed-order matching would be important.
One can see that the N$^3$LL result has a slightly reduced scale uncertainty compared to the NNLL$'$ one. The reduction is most significant in the small-$\tau$ region, where resummation effects are expected to be important. The N$^3$LL$'$ result further reduces the scale uncertainty in the intermediate $\tau$ region, with the three-loop hard, jet and soft functions included. However, we observe an unusual increase of scale uncertainty in the small $\tau$ region, as is clear from the right plot of Fig.~\ref{figure:hggn3ll}. It can be seen that the two bands even do not overlap below $\tau \sim 0.03$. This fact can be traced to the unusually large constant term $c^S_{3g}$ of the three-loop soft function. It is instructive to show the soft function at its default scale $\mu_s=m_H/\bar{N}$, where $L_S=0$, for $c_{3g}^S = -45433$:
\begin{equation}
\tilde{s}^g(0,\mu_s) = 1 - 2.356 \, \alpha_s(\mu_s) + 1.617 \, \alpha_s^2(\mu_s) - 22.90 \, \alpha_s^3(\mu_s) + \cdots \, .
\end{equation}
For small $\tau$, one expects that the dominant contributions in the Laplace space come from the region where $\tau N \sim 1$. This means that, below $\tau \sim 0.03$, $\mu_s$ is typically only about a few GeVs, where $\alpha_s \sim 0.2$ is not so small. Therefore, the gluon soft function has a rather poor perturbative convergence if we take the fitted central value of $c_{3g}^S$. It is highly desired to calculate the exact value of $c_{3g}^S$ to settle down this issue: either its absolute value is in fact smaller, and the N$^3$LL$'$ result is already sufficient; or it is indeed that large, then one needs to have even higher order corrections for reliable predictions. Efforts towards this goal are being actively pursued in the literature \cite{Baranowski:2022khd, Chen:2022cvz}.

\begin{figure}[t!]
\begin{center}
\includegraphics[width=0.45\textwidth]{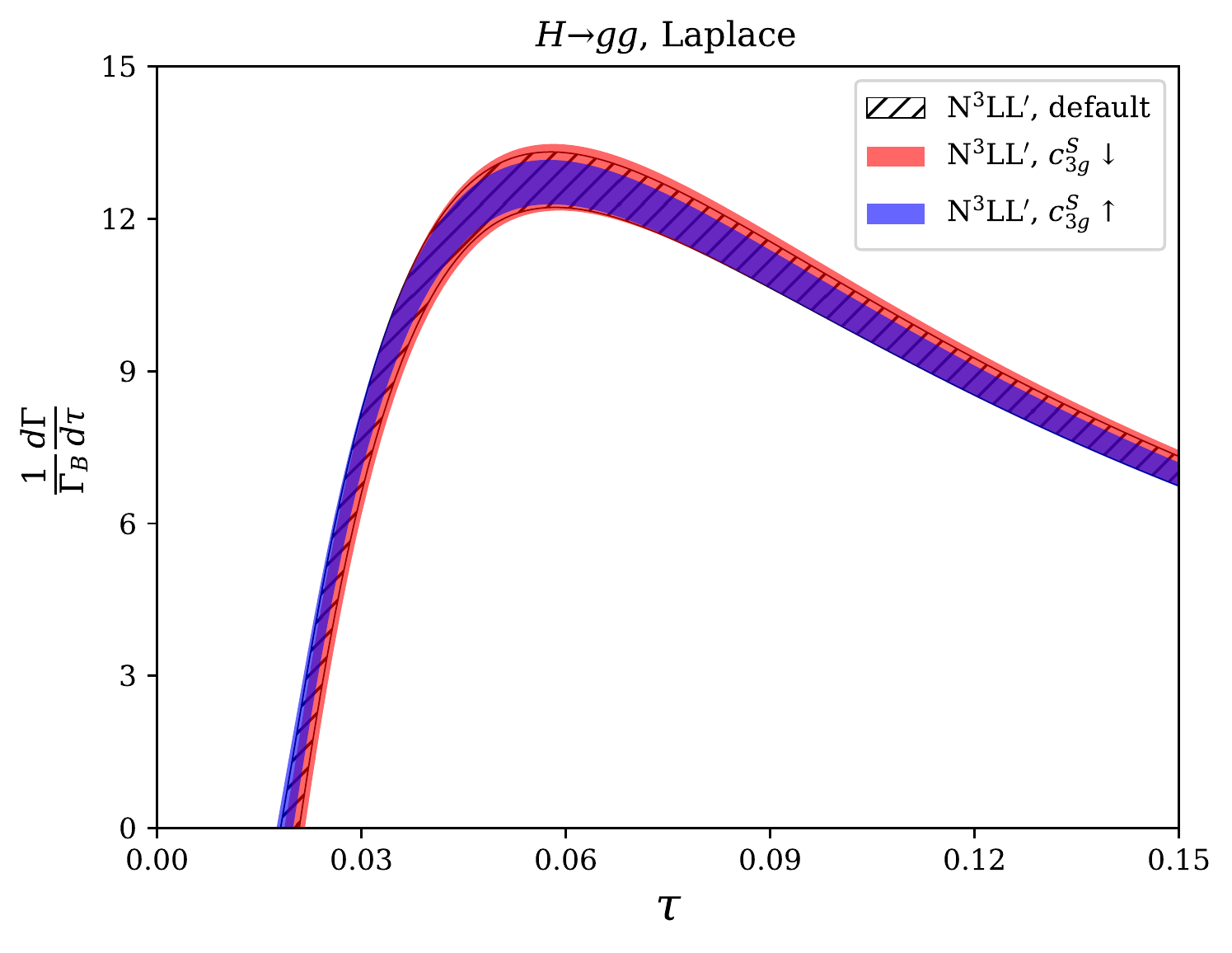}
\hspace{0.5cm}
\includegraphics[width=0.45\textwidth]{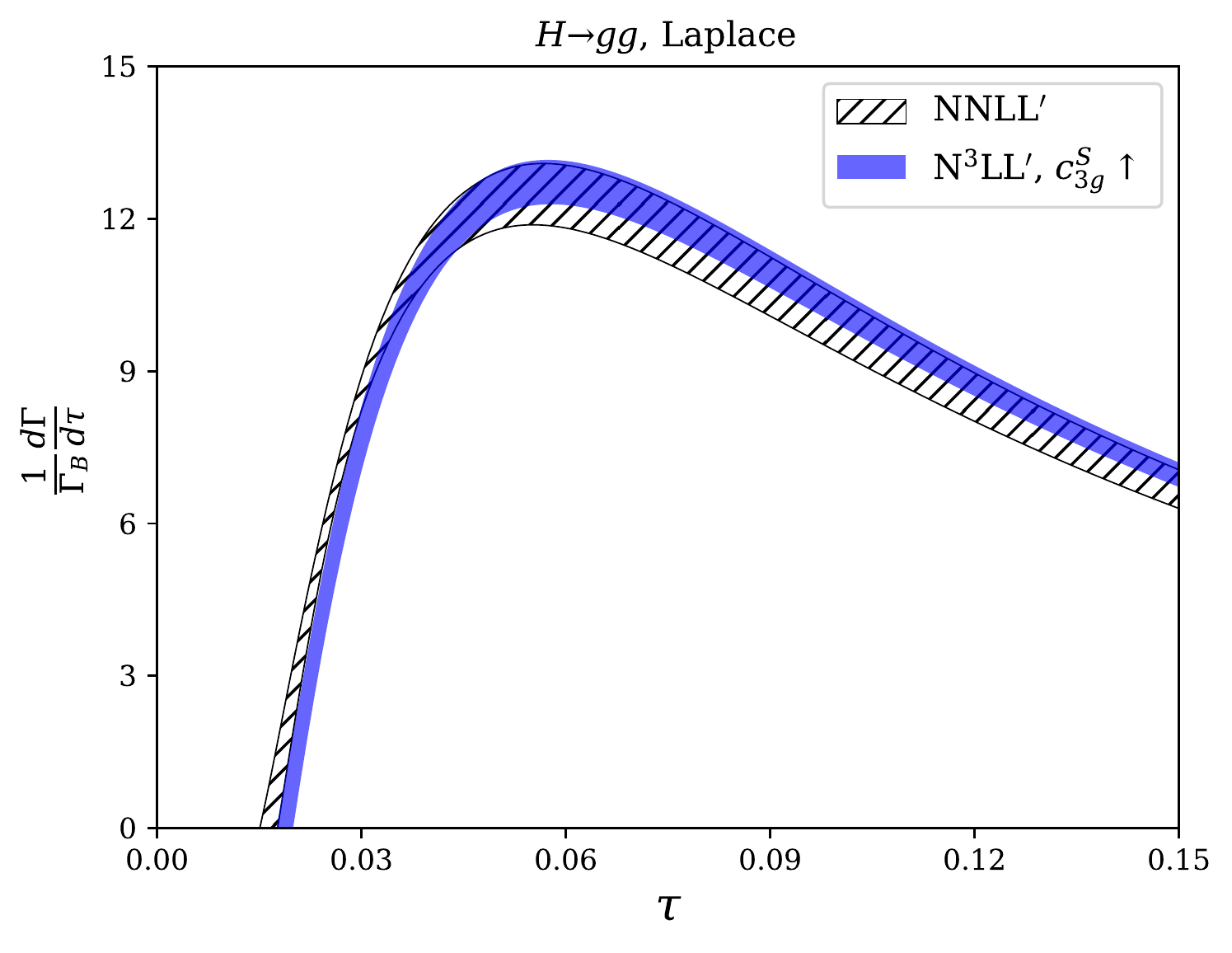}
\end{center}
\caption{\label{figure:hggn3llcs3}The effects of $c^S_{3g}$ on the N$^3$LL$'$ results in the gluon channel. Left: N$^3$LL$'$ results with 3 values of $c^S_{3g}$; Right: NNLL$'$ vs. N$^3$LL$'$ where $c^S_{3g}$ is taken to its ``upper'' value (with a smaller absolute value).}
\end{figure}

\begin{table}[t!]
	\centering
	\begin{tabular}{|c|c|c|c|c|c|}
		\hline
		$H \to gg$, N$^3$LL$'$ & $\mu_t$ & $\mu_h$ & $\mu_j$ & $\mu_s$ & $c_{3g}^S$
		\\ \hline
		max  & $12.128$ & $12.120 $ & $12.120$ & $13.077$& $12.219$
		\\ \hline
		min  & $12.106$ & $11.967$ & $12.063$ &$12.044$& $12.020$
		\\ \hline
	\end{tabular}
	\caption{Variations of the N$^3$LL$'$ differential decay rate at $\tau = 0.05$ induced by changing the scales and $c_{3g}^S$. The central value is $12.120$.\label{tab:N3LLp}}
\end{table}

To demonstrate the effects of different values of $c^S_{3g}$, we show in the left plot of Fig.~\ref{figure:hggn3llcs3} the resummed distributions for three values of $c^S_{3g}$: the default value $-45433$, the ``lower'' value $-56683$, and the ``upper'' value $-34183$. Note that since the fitted value of $c^S_{3g}$ is negative, the ``upper'' value has a smaller absolute value, and leads to a better perturbative convergence. Indeed, as can be seen from the plot, the result with the ``upper'' value exhibits a smaller scale uncertainty, especially in the small-$\tau$ region. It is also evident from the right plot of Fig.~\ref{figure:hggn3llcs3}, that the N$^3$LL$'$ band is better overlapped with the NNLL$'$ one with $c^S_{3g}$ taking the ``upper'' value. Finally for reference, we list in Table~\ref{tab:N3LLp} the variations of the N$^3$LL$'$ differential decay rate at $\tau = 0.05$ induced by changing the values of various scales as well as $c_{3g}^S$. It is clear that the main source of the scale uncertainty comes from the soft scale, as expected. It can also be seen that varying $c^S_{3g}$ has a larger effect than varying $\mu_t$, $\mu_h$ or $\mu_j$. All these emphasize again that we need a better understanding of the soft function at and beyond three loops.

\begin{figure}[t!]
\begin{center}
\includegraphics[width=0.45\textwidth]{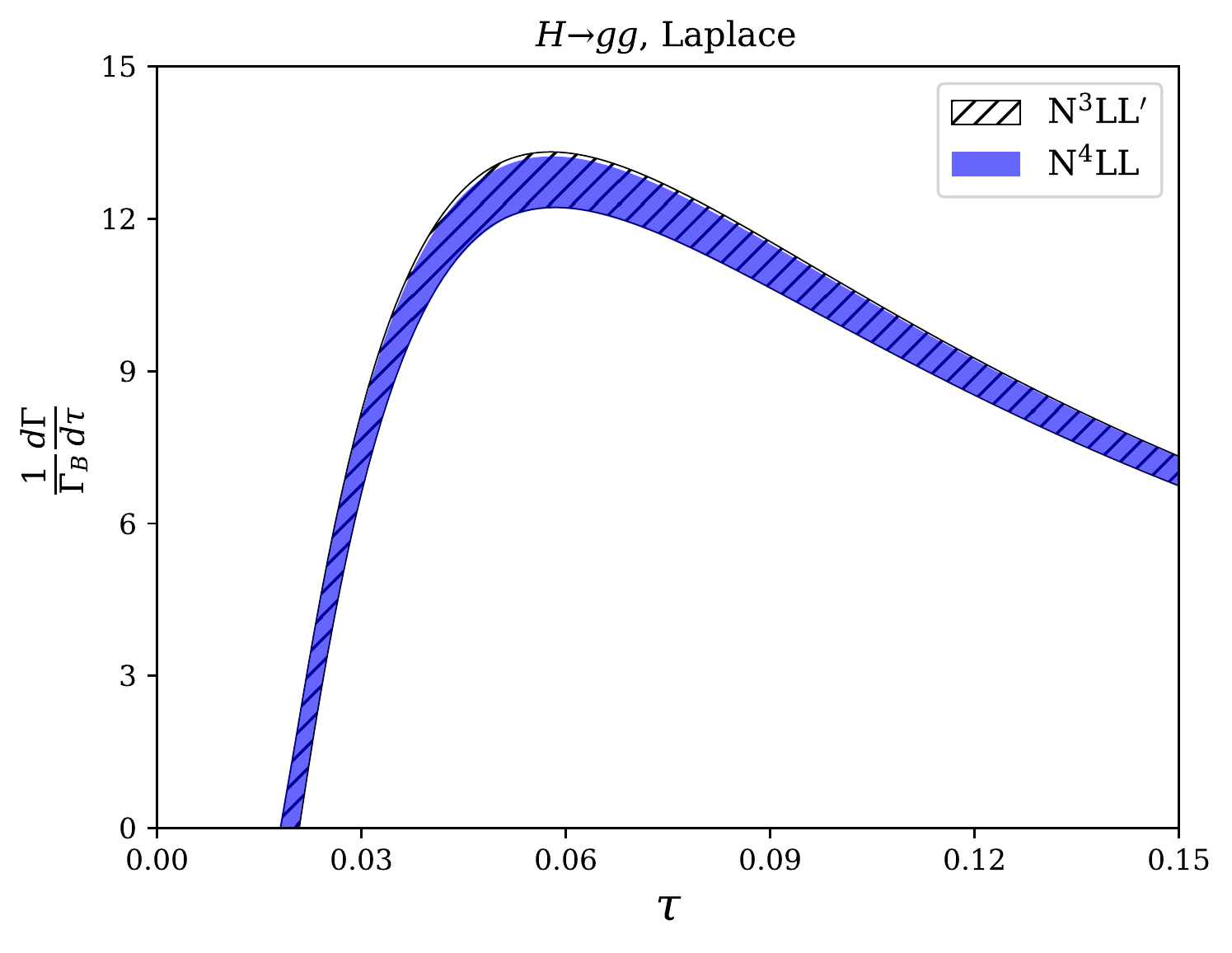}
\hspace{0.5cm}
\includegraphics[width=0.45\textwidth]{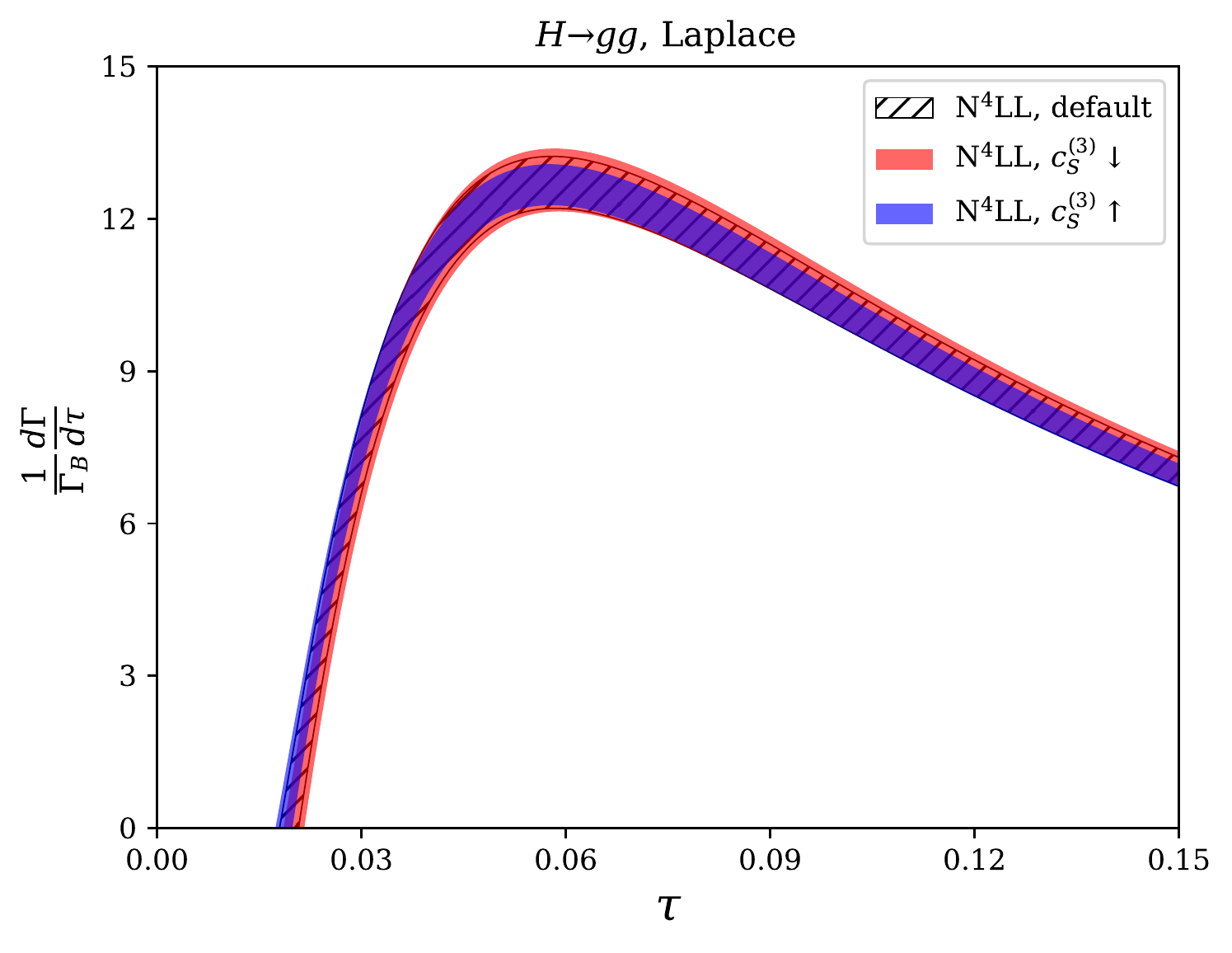}
\end{center}
\vspace{-1ex}
\caption{\label{figure:hggn4ll}The resummed thrust distributions in the gluon channel. Left: N$^3$LL$'$ vs N$^4$LL; Right: N$^4$LL results with 3 values of $c^S_{3g}$.}
\end{figure}

\begin{table}[t!]
	\centering
	\begin{tabular}{|c|c|c|c|c|c|c|}
		\hline
		  $H \to gg$, N$^4$LL & $\mu_t$ & $\mu_h$ & $\mu_j$ & $\mu_s$ & $c_{3g}^S$ & $\Gamma_{\text{cusp}}^{g(4)}$
		\\ \hline
		max & $12.089$ & $ 12.084$ & $ 12.084$ & $12.980$& $12.183$ & $12.085$
		\\ \hline
		min  & $12.073$ & $11.936$ & $12.025$ &$12.022$& $11.985$ & $12.083$
		\\ \hline
	\end{tabular}
	\caption{Variations of the N$^4$LL differential decay rate at $\tau = 0.05$ induced by changing the scales, $c_{3g}^S$ and the five-loop cusp anomalous dimension $\Gamma_{\text{cusp}}^{g(4)}$. The central value is $12.084$.\label{tab:N4LL}}
\end{table}

We now add another layer of resummation on top of N$^3$LL$'$, and present the results at N$^4$LL. The results are shown in Fig.~\ref{figure:hggn4ll}, with explicit numbers at $\tau=0.05$ given in Table~\ref{tab:N4LL}. We find that the additional order of resummation has a mild effect on the distribution, that is only clearly visible in the peak region. It is also evident that the five-loop cusp anomalous dimension does not have important impacts.

\subsection{The resummed thrust distributions in the quark-antiquark channel}

\begin{figure}[t!]
\begin{center}
\includegraphics[width=0.45\textwidth]{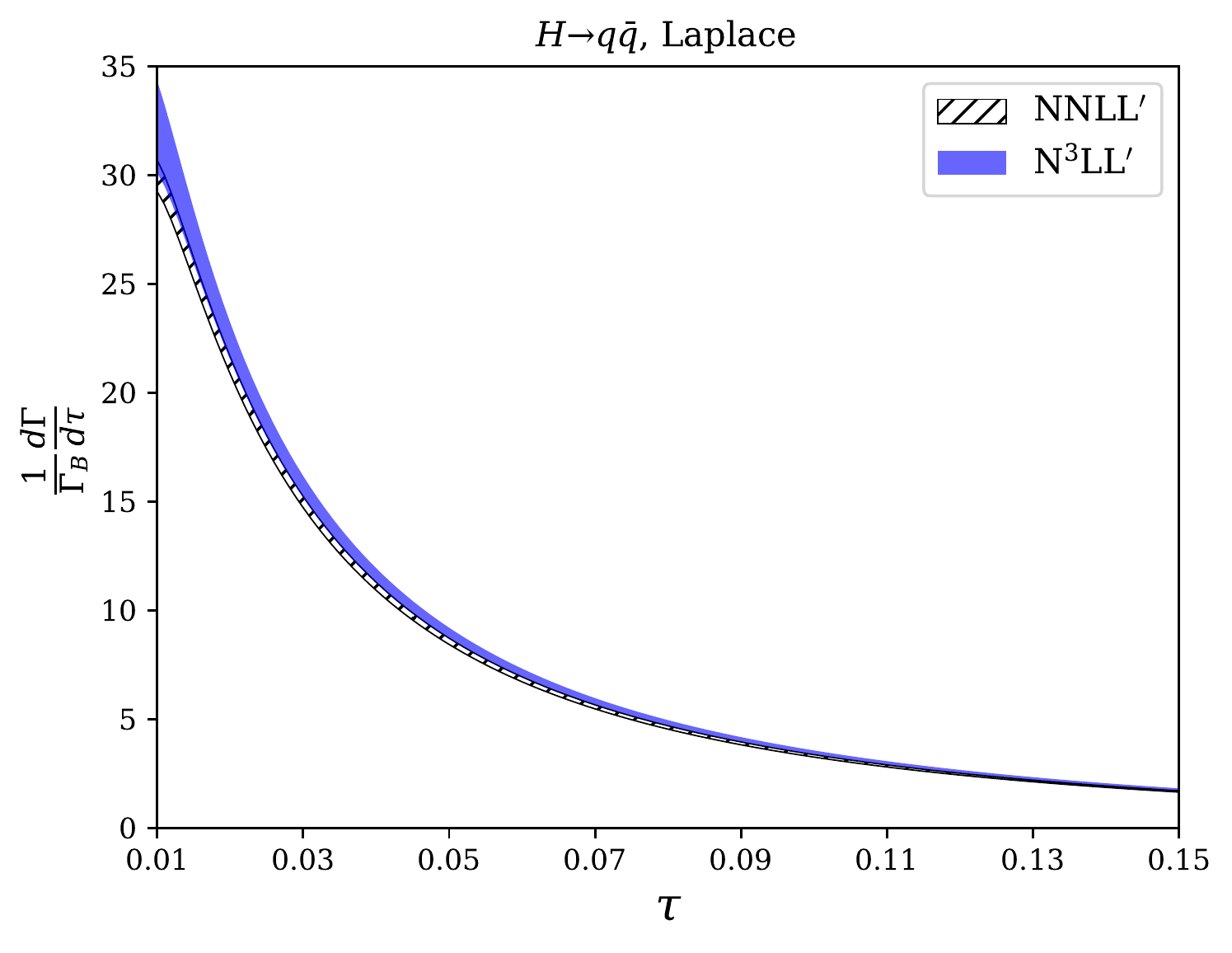}
\hspace{0.5cm}
\includegraphics[width=0.45\textwidth]{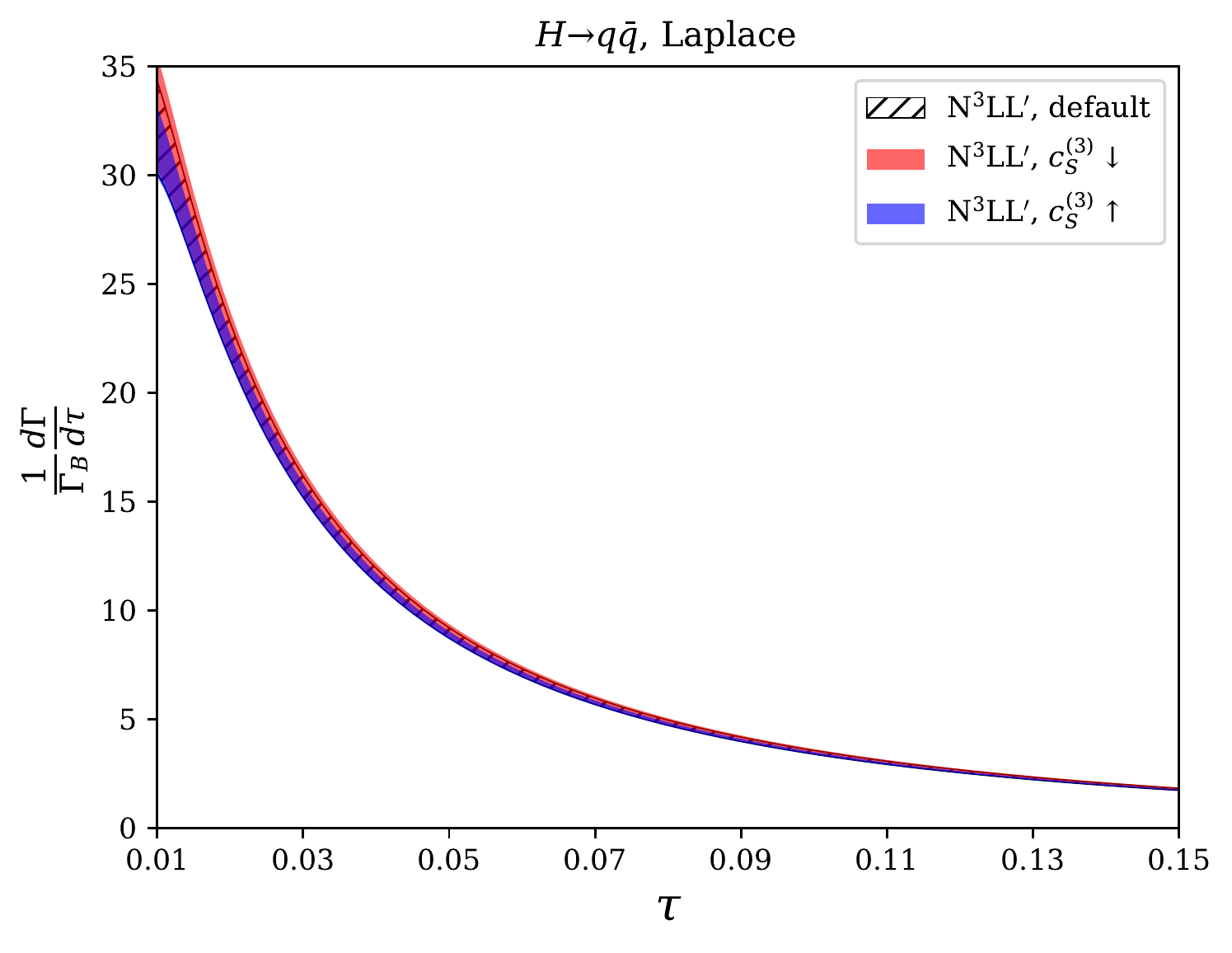}
\end{center}
\vspace{-1ex}
\caption{\label{figure:hqqn3llp}The resummed thrust distributions in the $q\bar{q}$ channel. Left: NNLL$'$ vs. N$^3$LL$'$; Right: N$^3$LL$'$ results with 3 values of $c_{3q}^S$.}
\end{figure}

We now briefly discuss the results in the quark-antiquark channel. In the left plot of Fig.~\ref{figure:hqqn3llp}, we compare the NNLL$'$ result against the N$^3$LL$'$ one, where the three-loop constant term $c^S_{3q}$ of the soft function is chosen at the default value. We again observe that the uncertainty band of N$^3$LL$'$ is broader than the NNLL$'$ one, especially at the lower end of the distribution. In the right plot of Fig.~\ref{figure:hqqn3llp}, we show the N$^3$LL$'$ distributions for 3 values of $c^S_{3q}$: the default value $-19988$, the ``upper'' value $-14988$ and the ``lower'' value $-24988$. As expected, the band becomes narrower for the ``upper'' value, where the absolute value of $c^S_{3q}$ is smaller, and the soft function has a perturbative convergence. Overall, the uncertainties of the resummed thrust distributions in the $q\bar{q}$ channel are significantly smaller than those in the gluon channel. This can be partly explained by the smaller color factor $C_F$ compared to $C_A$.

\section{Summary and outlook}
\label{sec:summary}

In this work, we extend the resummation for the thrust distribution in Higgs decays up to the fifth logarithmic order. A main conclusion that can be drawn from our results is that one needs the accurate values of the three-loop soft functions for reliable predictions in the small-$\tau$ region. This is especially true in the gluon channel, where the perturbative convergence of the soft function seems to be rather bad with a large three-loop constant term.

Once the three-loop soft functions become available, the ingredients collected in this work will allow for faithful numeric predictions at the N$^3$LL$'$ and N$^4$LL accuracies. Depending on the size of the three-loop constant, it is possible that one even needs the four-loop gluon soft function to reduce the scale uncertainties and obtain reliable predictions in the small-$\tau$ region.

\section*{Acknowledgments}

This work was supported in part by the National Natural Science Foundation of China under Grant No. 11975030 and 12147103, and the Fundamental Research Funds for the Central Universities.

\appendix

\section{Choice of scales in the momentum space}
\label{sec:app:momentum}

\begin{figure}[t!]
\centering
\includegraphics[width=0.45\textwidth]{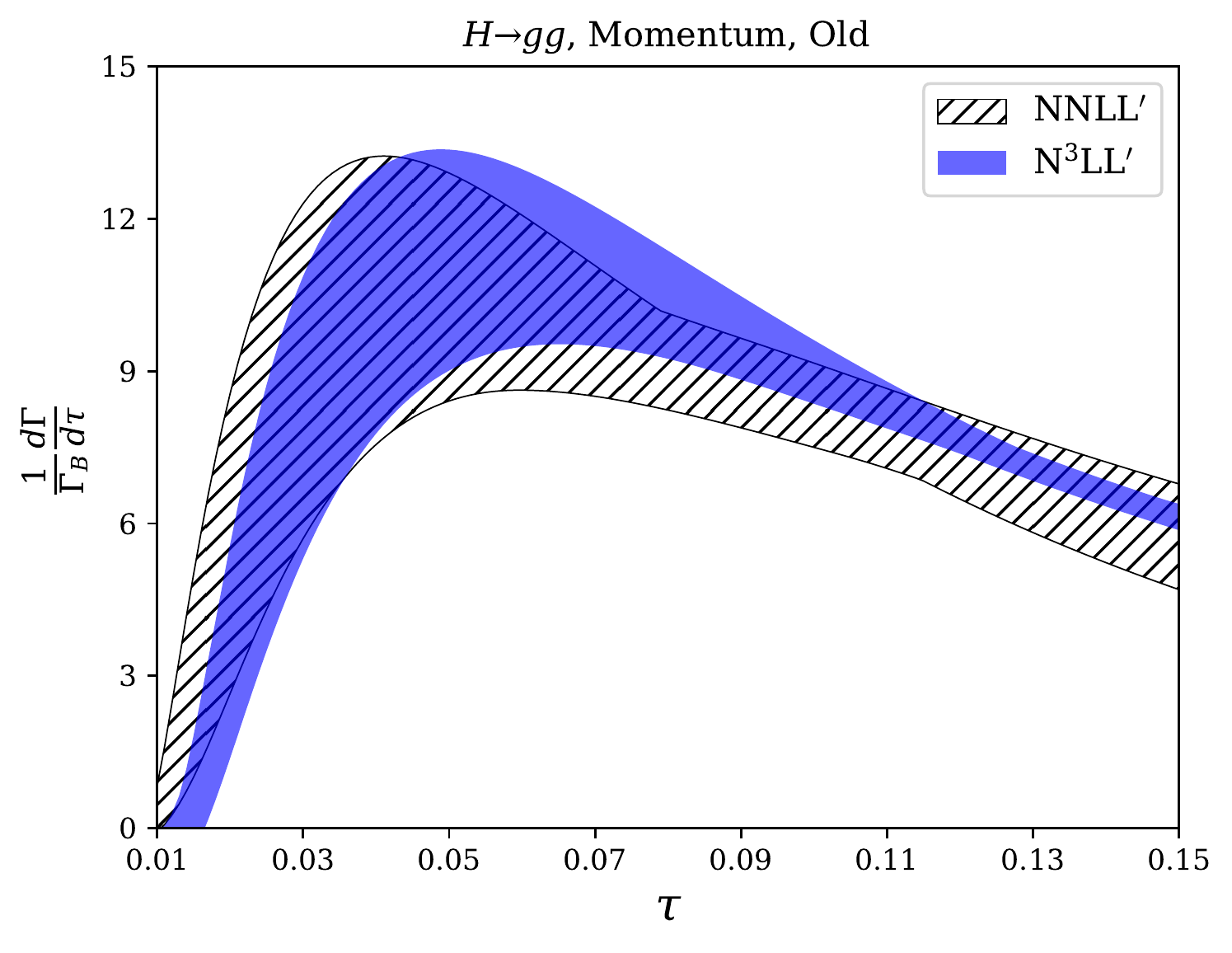}
\caption{The resummed thrust distributions at NNLL$'$ and N$^3$LL$'$ in the gluon channel with the scale choices \eqref{eq:scalesinmomentumspace} at the level of differential decay rates.\label{fig:momentum_nnllp_n3llp_old}}
\end{figure}

In the main text, we have chosen the jet and soft scales in the Laplace space, and performed the inverse Laplace transform numerically. A different approach is to set the jet and soft scales independent of the Laplace variable $N$. In this case, the inverse Laplace transform \eqref{eq:inverse_Laplace} can be carried out analytically \cite{Becher:2006nr, Becher:2006mr, Becher:2008cf}. For simplicity we only discuss the gluon channel in this Appendix. The result can be written as
\begin{align}\label{eq:resummed thrust distributions}
\frac{\diff\Gamma^g}{\diff\tau} &= \Gamma_B^{g}(\mu_h) \, U^g(\mu_t,\mu_h,\mu_j,\mu_s) \, \vert C_t (m_t,\mu_t) \vert^2 \, \vert C_S^g (m_H,\mu_h) \vert^2 \nonumber
\\
&\qquad \times \left[ \tilde{j}^g\left(  \ln \frac{\mu_s m_H}{\mu_j^2} + \partial_{\eta_g} , \mu_j \right) \right]^2 \tilde{s}^g(\partial_{\eta_g} , \mu_s) \left[ \frac{1}{\tau \, \Gamma(\eta_g)}
\left( \frac{\tau m_H}{\mu_s e^{ \gamma_E}} \right)^{\eta_g} \right] .
\end{align}
The common practice is then to choose $\mu_s$ and $\mu_j$ as a function of $\tau$, such that $\mu_j \sim \sqrt{\tau} \, m_H$ and $\mu_s \sim \tau m_H$ in the small-$\tau$ region. In this work we don't care about matching with the fixed-order results in the large-$\tau$ region (otherwise one needs to introduce ``profile scales'' as in \cite{Mo:2017gzp, Alioli:2020fzf}). Therefore we can adopt the simplest choices
\begin{equation}
\mu_t=e_t m_t \,, \quad \mu_h = e_h m_H \,, \quad \mu_j=e_j \sqrt{\tau} \, m_H \,, \quad \mu_s=e_s \tau m_H \, ,
\label{eq:scalesinmomentumspace}
\end{equation}
where the parameters $e_t$, $e_h$, $e_j$ and $e_s$ are set to $1$ by default, and are varied up and down by a factor of two to estimate the uncertainties. The numeric results with the above scale choices are shown in Fig.~\ref{fig:momentum_nnllp_n3llp_old}. We observe very large scale uncertainties in the small-$\tau$ region, much larger than those seen in Fig.~\ref{figure:hggn3ll}. As it turns out, these large uncertainties originate from the jet and soft scales, i.e., $e_j$ and $e_s$.

The prescription in Eq.~\eqref{eq:scalesinmomentumspace} actually has a subtle problem related exactly to the jet and soft scales. The formula \eqref{eq:resummed thrust distributions} is based on the solutions to the RG equations \eqref{eq:evolution equations in momentum space}. Taking the gluon jet function as an example, the RG equation is
\begin{align}
\frac{\diff}{\diff\ln\mu} J^g(p^2,\mu) &= \left[ -2 \Gamma^{g}_{\text{cusp}}(\alpha_s(\mu)) \ln\frac{p^2}{\mu^2} - 2\gamma^{g}_J(\alpha_s(\mu)) \right] J^{g}(p^2,\mu) \nonumber
\\
&\hspace{2cm} +2 \Gamma^{g}_{\text{cusp}}(\alpha_s(\mu)) \int^{p^2}_0 \diff q^2 \,  \frac{J^g(p^2,\mu)-J^g(q^2,\mu)}{p^2-q^2}\, .
\label{eq:RGE_jet}
\end{align}
And the solution reads \cite{Becher:2006nr, Becher:2006mr}
\begin{align}
J^g(p^2,\mu) = \exp \left[ -4S^g(\mu_j,\mu) + 2A^g_J(\mu_j,\mu) \right] \, \tilde{j}^g(\partial_{\eta},\mu_j) \left[ \frac{1}{p^2} \left( \frac{p^2}{\mu_j^2} \right)^{\eta} \right]_* \frac{e^{-\gamma_E \eta}}{\Gamma(\eta)} \, ,
\label{eq:sol_jet}
\end{align}
where $\eta = 2A^g_{\text{cusp}}(\mu_j,\mu)$, and the star-distribution is defined by
\begin{equation}
\int_0^{Q^2} dp^2 \left[ \frac{1}{p^2} \left( \frac{p^2}{\mu_j^2} \right)^{\eta} \right]_* f(p^2) = \int_0^{Q^2} dp^2 \, \frac{f(p^2)-f(0)}{p^2} \left( \frac{p^2}{\mu_j^2} \right)^\eta + \frac{f(0)}{\eta} \left( \frac{Q^2}{\mu^2} \right)^\eta \, .
\end{equation}
The solution is of course formally independent of $\mu_j$. However, at any finite order there is a residue dependence. Assuming that $\mu_j$ is independent of $p^2$, the above solution indeed satisfies the RG equation \eqref{eq:RGE_jet}, where $\mu_j$ is the same in both $J^g(p^2,\mu)$ and $J^g(q^2,\mu)$.
However, the scale choice in Eq.~\eqref{eq:scalesinmomentumspace} actually makes $\mu_j$ correlated with $p^2 \sim \tau m_H^2$. That is, $\mu_j^2 \sim p^2$ in $J^g(p^2,\mu)$, and $\mu_j^2 \sim q^2$ in $J^g(q^2,\mu)$. This immediately renders the convolution in Eq.~\eqref{eq:RGE_jet} ill-defined due to the singularity as $q^2 \to 0$.

One may of course ignore the problem with Eq.~\eqref{eq:RGE_jet} and insist on using Eq.~\eqref{eq:sol_jet} with $\mu_j \sim p^2$ as the resummed jet function. As long as one does not take $p^2 \to 0$ (where non-perturbative physics enters anyway), this does not pose any difficulty at face value. However, let us take a closer look. We set $\mu_j = e_j \sqrt{p^2}$ in Eq.~\eqref{eq:sol_jet}, and truncate the solution at NLL$'$ accuracy (that means two-loop $\Gamma_{\text{cusp}}^g$, one-loop $\gamma_J^g$ and one-loop $\tilde{j}^g$). We then expand the resummed jet function in terms of $\alpha_s(\mu)$. This gives
\begin{align}
J^{g,\text{NLL}'}\left(p^2,\mu; \mu_j=e_j \sqrt{p^2}\right) &= \frac{\alpha_s(\mu)}{4\pi} \frac{1}{p^2} \left( 12\ln\frac{p^2}{\mu^2} - \frac{23}{3} \right) + \left( \frac{\alpha_s(\mu)}{4\pi} \right)^2 \frac{1}{p^2} \nonumber
\\
&\hspace{-11em} \times \left[ 576 \ln^3(e_j) + 736 \ln^2(e_j) + \left( \frac{9364}{9} - 144\pi^2 \right) \ln(e_j) + 72\ln^3\left(\frac{p^2}{\mu^2}\right) + \cdots \right] + \mathcal{O}(\alpha_s^3) \, ,
\end{align}
where the ellipsis denotes further $e_j$-independent terms. We can see that the $e_j$-dependence cancels at order $\alpha_s^1$, as it should at the NLL$'$ accuracy. However, there exist rather high powers of $\ln(e_j)$ at order $\alpha_s^2$ (and beyond). While $\ln(e_j)$ is counted as a ``small'' logarithm, these high powers lead to large uncertainties of the resummed jet function when $e_j$ is varied between $1/2$ and $2$. The resummed soft function $S^g(k,\mu)$ with $\mu_s = e_s k$ exhibits the similar behavior. These explain the large uncertainty bands observed in Fig.~\ref{fig:momentum_nnllp_n3llp_old}. In practice, it is often argued that $e_j$ and $e_s$ are not independent and should be correlated. This can result in partial cancellations between the $\ln^n(e_j)$ and $\ln^n(e_s)$ terms, and lead to smaller uncertainty estimations.

The above behavior does not occur with the scale choices in Eq.~\eqref{eq:scales_Laplace}. We can do the same exercise with $\mu_j=e_j m_H/\sqrt{\bar{N}}$ in the Laplace space. The resummed Laplace-space jet function is given by
\begin{equation}
\tilde{j}^g(L_J,\mu) = \exp \left[ -4S^g(\mu_j,\mu) + 2A^g_J(\mu_j,\mu) \right] \left( \frac{m_H^2}{\bar{N} \, \mu_j^2} \right)^\eta \, \tilde{j}^g(L_J,\mu_j) \, .
\end{equation}
Again truncating at the NLL$'$ accuracy and expanding in terms of $\alpha_s(\mu)$, we can perform the inverse Laplace transform analytically to arrive at
\begin{align}
J^{g,\text{NLL}'}\left(p^2,\mu; \mu_j=e_j m_H/\sqrt{\bar{N}}\right) &= \frac{\alpha_s(\mu)}{4\pi} \frac{1}{p^2} \left( 12\ln\frac{p^2}{\mu^2} - \frac{23}{3} \right) \nonumber
\\
&+ \left( \frac{\alpha_s(\mu)}{4\pi} \right)^2 \frac{1}{p^2} \left[ 72\ln^3\left(\frac{p^2}{\mu^2}\right) + \cdots \right] + \mathcal{O}(\alpha_s^3) \, .
\end{align}
Evidently, all the $\ln^n(e_j)$ terms drop out at order $\alpha_s^2$. This explains the smaller uncertainty bands in Fig.~\ref{figure:hggn3ll} compared to Fig.~\ref{fig:momentum_nnllp_n3llp_old}.

There is an alternative way of choosing the jet and soft scales in the momentum space \cite{Almeida:2014uva,Bertolini:2017eui,Bell:2018gce}. We define the accumulative decay rate as
\begin{equation}
\hat{\Gamma}^g(\tau) \equiv \int_0^\tau \frac{\diff\Gamma^g}{\diff\tau'} \, d\tau' \, .
\end{equation}
In the above integrals, $\mu_j$ and $\mu_s$ are set to the same expressions as in Eq.~\eqref{eq:scalesinmomentumspace}, where $\tau$ is the upper bound of the integration. The jet and soft scales are hence independent of the integration variable $\tau'$. Before resummation, the factorization formula for the accumulative decay rates can be obtained by integrating Eq.~\eqref{eq: factorization formula} over $\tau$. The result reads
\begin{align}
\hat{\Gamma}^g(\tau) &= \Gamma_{B}^{g}(\mu) \, \vert C_t (m_t,\mu) \vert^2 \, \vert C_S^g (m_H,\mu) \vert^2 \int \diff p^2_{n} \, \diff p^2_{\bar{n}} \, \diff k \, \hat{J}_{n}^g(p^2_{n},\mu) \, \hat{J}_{\bar{n}}^g(p^2_{\bar{n}},\mu) \, S^g(k,\mu) \, ,
\end{align}
where the integration domain is determined by the constraints
\begin{equation}
\tau \geq \frac{p^2_{n}+p^2_{\bar{n}}}{m_H^2} + \frac{k}{m_H} \, , \quad p_n^2,p_{\bar{n}}^2,k \geq 0 \, .
\end{equation}
Now, since the jet scale $\mu_j$ is a function of $\tau$ and is independent of $p^2$, the problem with the convolution in Eq.~\eqref{eq:RGE_jet} is absent here.

\begin{figure}[t!]
\centering
\includegraphics[width=0.45\textwidth]{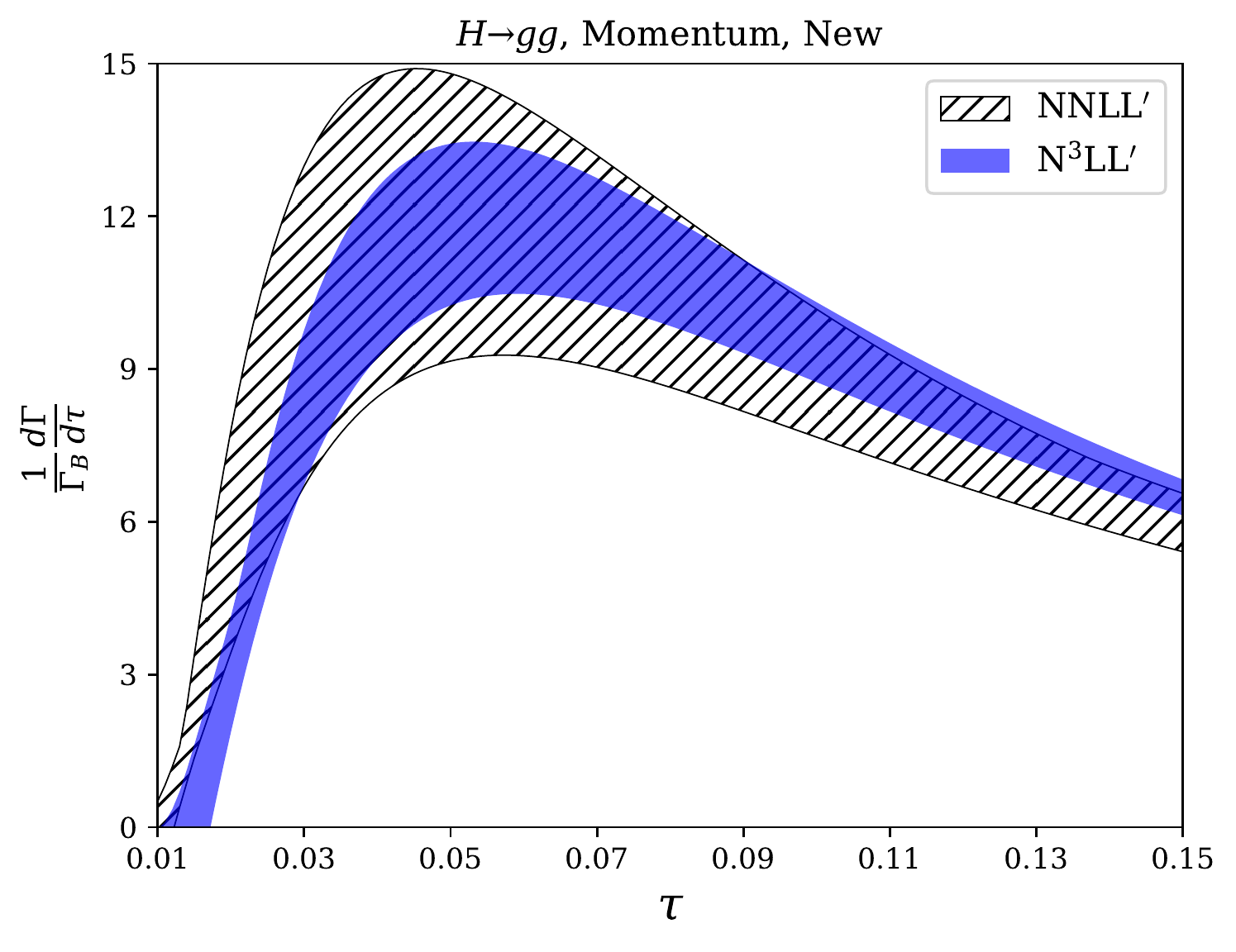}
\caption{The resummed thrust distributions at NNLL$'$ and N$^3$LL$'$ in the gluon channel with the scale choices \eqref{eq:scalesinmomentumspace} at the level of accumulative decay rates.\label{fig:momentum_nnllp_n3llp_new}}
\end{figure}

After resummation, the accumulative decay rate is
\begin{align}
\hat{\Gamma}^g(\tau) &= \Gamma_B^{g}(\mu_h) \, U^g(\mu_t,\mu_h,\mu_j,\mu_s) \, \vert C_t (m_t,\mu_t) \vert^2 \, \vert C_S^g (m_H,\mu_h) \vert^2 \nonumber
\\
&\qquad \times \left[ \tilde{j}^g\left(  \ln \frac{\mu_s m_H}{\mu_j^2} + \partial_{\eta_g} , \mu_j \right) \right]^2 \tilde{s}^g(\partial_{\eta_g} , \mu_s) \left[ \frac{1}{\Gamma(1+\eta_g)}
\left( \frac{\tau m_H}{\mu_s e^{ \gamma_E}} \right)^{\eta_g} \right] .
\label{eq:resummed_accumulative}
\end{align}
We set the jet and soft scales at the accumulative level following Eq.~\eqref{eq:scalesinmomentumspace}, i.e., $\mu_j = e_j \sqrt{\tau} m_H$ and $\mu_s = e_s \tau m_H$. We then take the derivative with respect to $\tau$ to obtain the resummed differential decay rate:
\begin{equation}
\frac{\diff\Gamma^g}{\diff\tau} = \frac{\diff}{\diff\tau} \hat{\Gamma}^g(\tau) \, .
\end{equation}
The numeric results with this new prescription are plotted in Fig.~\ref{fig:momentum_nnllp_n3llp_new}.

To compare the ``new'' way of scale setting at the accumulative level with the ``old'' way at the differential level, we can study the resummed integrated jet function
\begin{align}
\hat{J}^g(p^2,\mu) &= \int_0^{p^2} \diff q^2 \, J^g(q^2,\mu) \nonumber
\\
&= \exp \left[ -4S^g(\mu_j,\mu) + 2A^g_J(\mu_j,\mu) \right] \, \tilde{j}^g(\partial_{\eta},\mu_j) \left( \frac{p^2}{\mu_j^2} \right)^{\eta} \frac{e^{-\gamma_E \eta}}{\Gamma(1+\eta)} \, ,
\end{align}
with $\mu_j = e_j \sqrt{p^2}$. This is part of the resummed accumulative decay rate \eqref{eq:resummed_accumulative}. We still truncate at the NLL$'$ accuracy and expand in terms of $\alpha_s(\mu)$. We then take the derivative with respect to $p^2$ and arrive at
\begin{align}
\frac{\partial}{\partial p^2} \hat{J}^{g,\text{NLL}'}(p^2,\mu; \mu_j = e_j \sqrt{p^2}) &= \frac{\alpha_s(\mu)}{4\pi} \frac{1}{p^2} \left( 12 \ln \frac{p^2}{\mu^2} - \frac{23}{3} \right) \nonumber
\\
&+ \left( \frac{\alpha_s(\mu)}{4\pi} \right)^2 \frac{1}{p^2} \left[ 72 \ln^3 \left( \frac{p^2}{\mu^2} \right) + \cdots \right] + \mathcal{O}(\alpha_s^3) \, .
\end{align}
We see that the order $\alpha_s^2$ term is indeed independent of $e_j$.

\section{Fixed order ingredients}
\label{sec:app:fixed order ingredients}

In this Appendix we list the fixed-order ingredients appearing in the resummation formula \eqref{eq:Laplace-transformed resummed thrust distributions}.

The Wilson coefficient $C_t$ is known up to the four-loop order \cite{Spiridonov:1984br, Kramer:1996iq, Chetyrkin:1997un, Schroder:2005hy, Chetyrkin:2005ia,Liu:2015fxa}. For our purpose, we need its expression up to three loops, which is given by
\begin{align}
C_t(m_t,\mu) &= 1 + \frac{\alpha_s}{4\pi} \, 11 + \left( \frac{\alpha_s}{4\pi} \right)^2 \left[ L_t \left( 19 + \frac{16}{3} n_f \right) + \frac{2777}{18} - \frac{67}{6} n_f \right] \nonumber
\\
&+ \left( \frac{\alpha_s}{4\pi} \right)^3 \bigg[ L_t^2 \left( 209 + 46 n_f - \frac{32}{9} n_f^2 \right) + L_t \left( \frac{4834}{9} + \frac{2912}{27} n_f + \frac{77}{27} n_f^2 \right) \nonumber
\\
&\hspace{4em} -\frac{2761331}{648} + \frac{897943\zeta_3}{144}  + \left( \frac{58723}{324} - \frac{110779\zeta_3}{216} \right) n_f - \frac{6865}{486} n_f^2 \bigg] \, ,
\end{align}
where $L_t=\ln(\mu^2/m_t^2)$, and we have explicitly set the number of colors $N_c=3$ to shorten the expression.

The hard Wilson coefficients $C^{q,g}_S$ are expanded as
\begin{equation}
C_S^{q,g}(m_H,\mu) = \sum_{n=0} \left( \frac{\alpha_s}{4\pi} \right)^n C_S^{q,g(n)} \, ,
\end{equation}
where the gluon channel coefficients up to three-loop order are given by \cite{Gehrmann:2010ue}
\begin{align}
C_S^{g(0)} &= 1 \, , \nonumber
\\
C_S^{g(1)} &= \left( \frac{\pi^2}{6} - L^2 \right) C_A \, , \nonumber
\\
C_S^{g(2)} &= n_f C_A \left[-\frac{2 L^3}{9}+\frac{10 L^2}{9}+\left(\frac{52}{27}+\frac{2 \pi ^2}{9}\right) L-\frac{46 \zeta_3}{9}-\frac{5 \pi ^2}{18}-\frac{916}{81}\right]\nonumber
\\
&+ C_A^2 \left[\frac{L^4}{2}+\frac{11 L^3}{9}+\left(\frac{\pi ^2}{6}-\frac{67}{9}\right) L^2+L \left(-2 \zeta_3-\frac{11 \pi ^2}{9}+\frac{80}{27}\right)-\frac{143 \zeta_3}{9}\right.\nonumber\\
  &\left. +\frac{\pi ^4}{72}+\frac{67 \pi ^2}{36}+\frac{5105}{162}\right] + n_f C_F \left(2 L+8 \zeta_3-\frac{67}{6}\right),\nonumber
\\
C_S^{g(3)} &= n_f  C_A C_F \left[-\frac{8 L^3}{3}+L^2 (13-16 \zeta_3)+L \left(-\frac{376 \zeta_3}{9}+\frac{4 \pi ^4}{45}+\pi ^2+\frac{3833}{54}\right)\right.\nonumber\\
   &\left.+\frac{32 \pi ^2 \zeta_3}{9}+\frac{14564 \zeta_3}{81}+\frac{608 \zeta_5}{9}-\frac{34 \pi ^2}{27}-\frac{16 \pi ^4}{405}-\frac{341219}{972}\right]\nonumber\\
   &+ n_f^2 C_A \left[-\frac{2 L^4}{27}+\frac{40 L^3}{81}+\left(\frac{116}{81}+\frac{4 \pi ^2}{27}\right) L^2+L \left(-\frac{128 \zeta_3}{27}-\frac{40 \pi ^2}{81}-\frac{14057}{729}\right)\right.\nonumber\\
   &\left.+\frac{4576 \zeta_3}{243}+\frac{\pi ^4}{243}+\frac{2 \pi ^2}{27}+\frac{611401}{13122}\right]+ n_f C_A^2 \left[\frac{2 L^5}{9}-\frac{8 L^4}{27}+\left(-\frac{734}{81}-\frac{\pi ^2}{9}\right) L^3\right.\nonumber\\
   &\left.+L^2 \left(\frac{118 \zeta_3}{9}+\frac{377}{27}-\frac{103 \pi ^2}{54}\right)+L \left(\frac{28 \zeta_3}{9}+\frac{1910 \pi ^2}{243}-\frac{4 \pi ^4}{15}+\frac{133036}{729}\right)+\frac{428 \zeta_5}{9}\right.\nonumber\\
   &\left.-\frac{41 \pi ^2 \zeta_3}{27}-\frac{460 \zeta_3}{81}+\frac{73 \pi ^4}{1620}-\frac{14189 \pi ^2}{4374}-\frac{3765007}{6561}\right]+C_A^3 \left[-\frac{L^6}{6}-\frac{11 L^5}{9}\right.\nonumber\\
   &\left.+\left(\frac{281}{54}-\frac{\pi ^2}{4}\right) L^4+L^3 \left(2 \zeta_3+\frac{11 \pi ^2}{18}+\frac{1540}{81}\right)+L^2 \left(\frac{143 \zeta_3}{9}+\frac{685 \pi ^2}{108}-\frac{6740}{81}-\frac{73 \pi ^4}{360}\right)\right.\nonumber\\
   &\left. +L \left(\frac{17 \pi ^2 \zeta_3}{9}+\frac{2048 \zeta_3}{27}+16 \zeta_5+\frac{44 \pi ^4}{45}-\frac{6710 \pi ^2}{243}-\frac{373975}{1458}\right)+\frac{2222 \zeta_5}{9}\right.\nonumber\\
   &\left.-\frac{104 \zeta_3^2}{9}-\frac{605 \pi ^2 \zeta_3}{54}-\frac{152716 \zeta_3}{243}+\frac{105617 \pi ^2}{4374}-\frac{1939 \pi ^4}{9720}+\frac{29639273}{26244}-\frac{24389 \pi ^6}{408240}\right]\nonumber\\
   &+ n_f^2 C_F \left[\frac{4 L^2}{3}+L \left(\frac{32 \zeta_3}{3}-\frac{52}{3}\right)-\frac{112 \zeta_3}{3}-\frac{10 \pi ^2}{27}+\frac{4481}{81}-\frac{4 \pi ^4}{405}\right]\nonumber\\
   &+ n_f C_F^2 \left[-2 L+\frac{296 \zeta_3}{3}-160 \zeta_5+\frac{304}{9}\right],
\label{eq:hardfunctionExplicitHgg}
\end{align}
and the quark channel coefficients are given by \cite{Gehrmann:2014vha}
\begin{align}
C_S^{q(0)} &= 1 \, , \nonumber
\\
C_S^{q(1)} &= \frac{1}{6} \left(-6 L^2+\pi ^2-12\right) C_F \, , \nonumber
\\
C_S^{q(2)} &= C_F^2\left[\frac{L^4}{2}+\left(2-\frac{\pi ^2}{6}\right) L^2+6 (4 L-5) \zeta_3-2 \pi ^2 L+\frac{7 \pi ^2}{3}-\frac{83 \pi ^4}{360}+6\right]\nonumber\\
&+C_F\left[ \frac{11 L^3 C_A}{9}+\frac{1}{3} \pi ^2 L^2 C_A-\frac{67 L^2 C_A}{9}-26 L \zeta_3 C_A+\frac{11}{9} \pi ^2 L C_A+\frac{242 L C_A}{27}+\right.\nonumber\\
&\left.\frac{151 \zeta_3 C_A}{9}+\frac{11 \pi ^4 C_A}{45}-\frac{467 C_A}{81}-\frac{103 \pi ^2 C_A}{108}-\frac{10 L^3}{9}+\frac{50 L^2}{9}-\frac{10 \pi ^2 L}{9}-\frac{280 L}{27}+\frac{10 \zeta_3}{9}\right.\nonumber\\
&\left.+\frac{25 \pi ^2}{54}+\frac{1000}{81}\right],\nonumber
\\
C_S^{q(3)} &= C_F^3\left[ -\frac{L^6}{6}+\frac{\pi ^2 L^4}{12}-L^4-24 L^3 \zeta_3+2 \pi ^2 L^3+30 L^2 \zeta_3+\frac{83 \pi ^4 L^2}{360}-\frac{7 \pi ^2 L^2}{3}-6 L^2\right.\nonumber\\
&\left.-240 L \zeta_5-\frac{4}{3} \pi ^2 L \zeta_3+20 L \zeta_3+\frac{19 \pi ^4 L}{15}+7 \pi ^2 L-50 L+16 \zeta_3^2+\frac{89 \pi ^2 \zeta_3}{3}\right.\nonumber\\
&\left.-654 \zeta_3+424 \zeta_5+\frac{37729 \pi ^6}{136080}+\frac{575}{3}-\frac{353 \pi ^2}{18}-\frac{77 \pi ^4}{36}\right]
+C_F^2C_A\left[ -\frac{11 L^5}{9}-\frac{\pi ^2 L^4}{3}\right.\nonumber\\
&\left.+\frac{67 L^4}{9}+26 L^3 \zeta_3-\frac{308 L^3}{27}-\frac{55 \pi ^2 L^3}{54}-\frac{943 L^2 \zeta_3}{9}+\frac{689 \pi ^2 L^2}{108}+\frac{1673 L^2}{81}-\frac{17 \pi ^4 L^2}{90}\right.\nonumber\\
&\left.-\frac{5}{3} \pi ^2 L \zeta_3+\frac{1660 L \zeta_3}{3}+120 L \zeta_5+\frac{\pi ^4 L}{6}+\frac{614 L}{27}-\frac{3506 \pi ^2 L}{81}+\frac{296 \zeta_3^2}{3}-\frac{1676 \zeta_5}{9}\right.\nonumber\\
&\left.-\frac{4820 \zeta_3}{27}-\frac{3049 \pi ^2 \zeta_3}{54}+\frac{31819 \pi ^2}{486}-\frac{9335}{81}-\frac{893 \pi ^4}{9720}-\frac{3169 \pi ^6}{17010}\right]+C_F^2 \left[ \frac{10 L^5}{9}-\frac{50 L^4}{9}\right.\nonumber\\
&\left.+\frac{25 \pi ^2 L^3}{27}+\frac{250 L^3}{27}+\frac{350 L^2 \zeta_3}{9}-\frac{335 \pi ^2 L^2}{54}+\frac{3625 L^2}{162}-\frac{4160 L \zeta_3}{9}+\frac{7 \pi ^4 L}{9}+\frac{2200 \pi ^2 L}{81}\right.\nonumber\\
&\left.-\frac{7075 L}{54}+\frac{59980 \zeta_3}{81}-\frac{2080 \zeta_5}{9}-\frac{95 \pi ^2 \zeta_3}{27}-\frac{305 \pi ^4}{972}-\frac{30655 \pi ^2}{972}+\frac{179375}{972}\right] \nonumber\\
& +C_FC_A^2\left[ -\frac{121 L^4}{54}-\frac{22 \pi ^2 L^3}{27}+\frac{1780 L^3}{81}+88 L^2 \zeta_3+\frac{13 \pi ^2 L^2}{27}-\frac{11 \pi ^4 L^2}{45}-\frac{11939 L^2}{162}\right.\nonumber\\
&\left.+\frac{44}{9} \pi ^2 L \zeta_3+136 L \zeta_5-\frac{13900 L \zeta_3}{27}+\frac{4822 \pi ^2 L}{243}-\frac{47 \pi ^4 L}{54}+\frac{10289 L}{1458}+\frac{163 \pi ^2 \zeta_3}{9}\right.\nonumber\\
&\left.+\frac{107648 \zeta_3}{243} +\frac{106 \zeta_5}{9}-\frac{1136 \zeta_3^2}{9}+\frac{10093 \pi ^4}{4860}-\frac{769 \pi ^6}{5103}-\frac{264515 \pi ^2}{8748}+\frac{5964431}{26244}\right]\nonumber\\
&+C_AC_F\left[\frac{110 L^4}{27}+\frac{20 \pi ^2 L^3}{27}-\frac{2890 L^3}{81}-40 L^2 \zeta_3+\frac{40 \pi ^2 L^2}{9}+\frac{8635 L^2}{81}+\frac{3620 L \zeta_3}{9}\right.\nonumber\\
&\left.+\frac{11 \pi ^4 L}{27}-\frac{8180 \pi ^2 L}{243}-\frac{37495 L}{729}+\frac{10 \pi ^2 \zeta_3}{9}-\frac{20 \zeta_5}{3}-\frac{14300 \zeta_3}{27}+\frac{166295 \pi ^2}{4374}\right.\nonumber\\
&\left.-\frac{119 \pi ^4}{243}-\frac{2609875}{13122}\right]+C_F\left[ -\frac{50 L^4}{27}+\frac{1000 L^3}{81}-\frac{100 \pi ^2 L^2}{27}-\frac{2500 L^2}{81}+\frac{400 L \zeta_3}{27}\right.\nonumber\\
&\left.+\frac{1000 \pi ^2 L}{81}+\frac{23200 L}{729}-\frac{5000 \zeta_3}{243}-\frac{235 \pi ^4}{243}-\frac{2650 \pi ^2}{243}+\frac{51800}{6561}\right] ,
\label{eq:hardfunctionExplicitHqq}
\end{align}
where
$L=\ln \left[ (-m_H^2-i \epsilon)/\mu^2\right] $. Although not used in this paper, the fourth-order coefficients can already be extracted from the form factors calculated in Ref.~\cite{Chakraborty:2022yan} and  Ref.~\cite{Lee:2022nhh}.

The jet functions are expanded as
\begin{equation}
\tilde{j}^{q,g}(L_J,\mu) = \sum_{n=0}\left(\frac{\alpha_s}{4\pi}\right)^n \tilde{j}^{q,g(n)}(L_J) \, .
\end{equation}
The expansion coefficients for the quark jet function are \cite{Becher:2008cf, Bruser:2018rad}
\begin{align}
\tilde{j}^{q(1)}(L_J) &= C_F\left( 2 L_J^2-3 L_J-\frac{2 \pi ^2}{3}+7\right), \nonumber
\\
\tilde{j}^{q(2)}(L_J) &= C_Fn_f \left[ \frac{4}{9} L_J^3 - \frac{29}{9} L_J^2 + \left( \frac{247}{27} - \frac{2\pi^2}{9} \right) L_J + \frac{13\pi^2}{18} - \frac{4057}{324} \right]
\nonumber
\\
&\hspace{-3em} + C_FC_A \left[ -\frac{22}{9} L_J^3 + \left( \frac{367}{18} - \frac{2\pi^2}{3} \right) L_J^2 + \left( 40 \zeta_3 + \frac{11\pi^2}{9} - \frac{3155}{54} \right) L_J - 18 \zeta_3 - \frac{37\pi^4}{180} \right. \nonumber
\\
&\hspace{-3em} \left. -\frac{155\pi^2}{36} + \frac{53129}{648} \right] + C_F^2 \left[ 2 L_J^4 - 6 L_J^3 + \left( \frac{37}{2} - \frac{4\pi^2}{3} \right) L_J^2 + \left( 4\pi^2 - 24 \zeta_3 - \frac{45}{2} \right) L_J
\right.
\nonumber
\\
&\hspace{-3em} \left. - 6 \zeta_3 + \frac{61\pi^4}{90} - \frac{97\pi^2}{12} + \frac{205}{8} \right] \, ,
\nonumber
\\
\tilde{j}^{q(3)}(L_J) &= C_F n_f^2 \Bigg[ \frac{4}{27} L_J^4 - \frac{116}{81} L_J^3 + \left( \frac{470}{81} - \frac{4\pi^2}{27} \right) L_J^2 + \bigg( \frac{58\pi^2}{81} - \frac{8714}{729} - \frac{64}{27} \zeta_3 \bigg) L_J \Bigg]
\nonumber
\\
&\hspace{-3em} + C_FC_An_f \Bigg[ -\frac{44}{27} L_J^4 + \left( \frac{1552}{81} - \frac{8\pi^2}{27} \right) L_J^3 + \left( \frac{28\pi^2}{9} - \frac{7531}{81} + 8 \zeta_3 \right) L_J^2 +\bigg( \frac{32\pi^4}{135}
\nonumber
\\
&\hspace{-3em} - \frac{1976\zeta_3}{27} - \frac{2632\pi^2}{243} + \frac{160906}{729} \bigg) L_J \Bigg]
+ C_FC_A^2 \Bigg[  \frac{121}{27} L_J^4 + \bigg( \frac{44\pi^2}{27} - \frac{4649}{81} \bigg) L_J^3+ \bigg( \frac{22\pi^4}{45}
\nonumber
\\
&\hspace{-3em} -132\zeta_3 - \frac{389\pi^2}{27} + \frac{50689}{162} \bigg) L_J^2 + \bigg( \frac{18179\pi^2}{486} - \frac{53\pi^4}{135} - \frac{599375}{729} - 232 \zeta_5 - \frac{88\pi^2\zeta_3}{9}
\nonumber
\\
&\hspace{-3em} + \frac{6688 \zeta_3}{9} \bigg) L_J \Bigg] + C_F^2 n_f \Bigg[ \frac{8}{9} L_J^5 - \frac{70}{9} L_J^4 + \bigg( \frac{875}{27} - \frac{20\pi^2}{27} \bigg) L_J^3 + \bigg( \frac{151\pi^2}{27} - \frac{15775}{162} \bigg) L_J^2
\nonumber
\\
&\hspace{-3em} + \bigg(\frac{32\zeta_3}{9} + \frac{4\pi^4}{27} - \frac{2833\pi^2}{162} + \frac{7325}{36} \bigg) L_J \Bigg]
+ C_F^2C_A \Bigg[ -\frac{44}{9} L_J^5 + \bigg( \frac{433}{9} - \frac{4\pi^2}{3} \bigg) L_J^4
\nonumber
\\
&\hspace{-3em} + \bigg( \frac{164\pi^2}{27} - \frac{10537}{54} + 80 \zeta_3 \bigg) L_J^3 + \bigg( -68 \zeta_3 + \frac{\pi^4}{30} - \frac{2045\pi^2}{54} + \frac{157943}{324} \bigg) L_J^2
\nonumber
\\
&\hspace{-3em} + \bigg( \frac{290\zeta_3}{3} - 120 \zeta_5 - \frac{88\pi^2\zeta_3}{3} - \frac{923\pi^4}{540} + \frac{35075\pi^2}{324} - \frac{151405}{216} \bigg) L_J \Bigg] + C_F^3 \Bigg[  \frac{4}{3} L_J^6 - 6 L_J^5
\nonumber
\\
&\hspace{-3em} + \bigg( 23 - \frac{4\pi^2}{3} \bigg) L_J^4 + \bigg( 8\pi^2 - \frac{99}{2} - 48\zeta_3 \bigg) L_J^3 + \bigg( 60 \zeta_3 + \frac{61\pi^4}{45} - \frac{151\pi^2}{6} + \frac{349}{4} \bigg) L_J^2
\nonumber
\\
&\hspace{-3em} + \bigg( 240 \zeta_5 + \frac{64\pi^2\zeta_3}{3} - 218 \zeta_3 - \frac{149\pi^4}{30} + \frac{145\pi^2}{4} - \frac{815}{8} \bigg) L_J\Bigg] + c_{3q}^J \, ,
\end{align}
The scale-independent constant term $ c^J_{3q}$ is given by \cite{Bruser:2018rad}
\begin{align}
c^J_{3q} &= 25.06777873 C_F^3+ 32.81169125  C_A   C_F^2 - 0.7795843561 C_A^2 C_F- 31.65196210C_AC_Fn_fT_F \nonumber
\\
&-61.78995095 C_F^2n_fT_F + 28.49157341C_Fn_f^2T_F^2 \, .
\end{align}
The expansion coefficients for the gluon jet function are \cite{Becher:2009th, Banerjee:2018ozf}
\begin{align}
\tilde{j}^{g(1)}(L_J) &= C_A \left( 2L_J^2 - \frac{11}{3} L_J + \frac{67}{9} - \frac{2\pi^2}{3}\right) + n_f \left( \frac{2}{3} L_J - \frac{10}{9} \right) , \nonumber
\\
\tilde{j}^{g(2)}(L_J) &= n_f^2 \left( \frac{4}{9} L_J^2 - \frac{40}{27} L_J - \frac{2\pi^2}{27} + \frac{100}{81} \right)
+ C_Fn_f \left( 2L_J + 8\zeta_3 - \frac{55}{6} \right) + C_An_f \Bigg[ \frac{16}{9} L_J^3
\nonumber
\\
&\hspace{-3em} - \frac{28}{3} L_J^2 + \bigg( \frac{224}{9} - \frac{10\pi^2}{9} \bigg) L_J - \frac{8\zeta_3}{3} + \frac{67\pi^2}{27} - \frac{760}{27} \Bigg]
+ C_A^2 \Bigg[ 2 L_J^4 - \frac{88}{9} L_J^3 + \bigg( \frac{389}{9} - 2\pi^2 \bigg) L_J^2
\nonumber
\\
&\hspace{-3em} + \bigg( \frac{55\pi^2}{9} + 16\zeta_3 - \frac{2570}{27} \bigg) L_J - \frac{88\zeta_3}{3} + \frac{17\pi^4}{36} - \frac{362\pi^2}{27} + \frac{20215}{162} \Bigg] \, ,
\nonumber
\\
\tilde{j}^{g(3)}(L_J) &= n_f^3 \Bigg[ \frac{8}{27} L_J^3 - \frac{40}{27} L_J^2 + \bigg( \frac{200}{81} - \frac{4\pi^2}{27} \bigg) L_J \Bigg] + C_F n_f^2 \Bigg[ \frac{10}{3} L_J^2 + \bigg( 16\zeta_3 - 24 \bigg) L_J \Bigg]
\nonumber
\\
&\hspace{-3em} - C_F^2n_f L_J + C_An_f^2 \Bigg[ \frac{4}{3} L_J^4 - \frac{292}{27} L_J^3 + \bigg( \frac{3326}{81} - \frac{4\pi^2}{3} \bigg) L_J^2 + \bigg( \frac{508\pi^2}{81} - \frac{116509}{1458} - \frac{256\zeta_3}{27} \bigg) L_J \Bigg]
\nonumber
\\
&\hspace{-3em} + C_AC_Fn_f \Bigg[ \frac{16}{3} L_J^3 + \big( 32\zeta_3 - 55 \big) L_J^2 + \bigg( -\frac{8\pi^4}{45} - \frac{10\pi^2}{3} + \frac{5599}{27} - \frac{1096\zeta_3}{9} \bigg) L_J \Bigg]
\nonumber
\\
&\hspace{-3em} + C_A^2n_f \Bigg[ \frac{20}{9} L_J^5 - \frac{64}{3} L_J^4 - \bigg( \frac{88\pi^2}{27} - \frac{3106}{27} \bigg) L_J^3 + \bigg( \frac{586\pi^2}{27} - \frac{8\zeta_3}{3}     -\frac{10067}{27} \bigg)  L_J^2
\nonumber
\\
&\hspace{-3em} + \bigg( \frac{449\pi^4}{270} - \frac{16831\pi^2}{243} + \frac{1052135}{1458} - \frac{1280\zeta_3}{27} \bigg) L_J \Bigg] + C_A^3 \Bigg[ \frac{4}{3} L_J^6 - \frac{110}{9}
L_J^5 + \bigg( 85 - \frac{8\pi^2}{3} \bigg) L_J^4
\nonumber
\\
&\hspace{-3em} + \bigg( \frac{484\pi^2}{27} - \frac{9623}{27} + 32 \zeta_3 \bigg) L_J^3 + \bigg( \frac{169\pi^4}{90} - \frac{484\zeta_3}{3} - \frac{2362\pi^2}{27} + \frac{85924}{81} \bigg) L_J^2
\nonumber
\\
&\hspace{-3em} + \bigg( -\frac{4411\pi^4}{540} + \frac{52678\pi^2}{243} - \frac{1448021}{729} - 112\zeta_5 - \frac{160\pi^2\zeta_3}{9} + \frac{6316\zeta_3}{9} \bigg) L_J \Bigg] + c^J_{3g} \, .
\end{align}
From Ref.~\cite{Banerjee:2018ozf}, we have $c^J_{3g} =647.7843434644$ for $n_f=5$.

The soft functions are expanded as
\begin{equation}
\tilde{s}_{q,g}(L_S,\mu)=\sum_{n=0}\left(\frac{\alpha_s}{4\pi}\right)^n \tilde{s}^{q,g(n)}(L_S) \, .
\end{equation}
The expansion coefficients for the quark soft function are \cite{Becher:2008cf, Kelley:2011ng}
\begin{align}
\tilde{s}^{q(1)}(L_S) &= C_F \left( -8L_S^2 - \pi^2 \right) ,
\nonumber
\\
\tilde{s}^{q(2)}(L_S) &= C_Fn_f \Bigg[ -\frac{32}{9} L_S^3 + \frac{80}{9} L_S^2 - \bigg( \frac{8\pi^2}{9} + \frac{224}{27} \bigg) L_S - \frac{52\zeta_3}{9} + \frac{77\pi^2}{27} + \frac{40}{81} \Bigg]
\nonumber
\\
&\hspace{-3em} + C_F C_A \Bigg[ \frac{176}{9} L_S^3 + \bigg( \frac{8\pi^2}{3} - \frac{536}{9} \bigg) L_S^2 + \bigg( \frac{44\pi^2}{9} - 56 \zeta_3 + \frac{1616}{27} \bigg) L_S + \frac{286\zeta_3}{9} + \frac{14\pi^4}{15}
\nonumber
\\
&\hspace{-3em} - \frac{871\pi^2}{54} - \frac{2140}{81} \Bigg]
+ C_F^2 \Bigg( 32 L_S^4 + 8\pi^2 L_S^2 + \frac{\pi^4}{2}\Bigg) \, ,
\nonumber
\\
\tilde{s}^{q(3)}(L_S) &= C_Fn_f^2 \Bigg[ -\frac{64}{27} L_S^4 + \frac{640}{81} L_S^3 - \bigg (\frac{32\pi^2}{27} + \frac{800}{81} \bigg) L_S^2 + \bigg( \frac{64\pi^2}{9} - \frac{3200}{729} - \frac{64\zeta_3}{9} \bigg) L_S \Bigg]
\nonumber
\\
&\hspace{-3em} + C_FC_An_f \Bigg[ \frac{704}{27} L_S^4 + \bigg( \frac{64\pi^2}{27} - \frac{9248}{81} \bigg) L_S^3 + \bigg( \frac{64\pi^2}{9} + \frac{16408}{81} \bigg) L_S^2 + \bigg( \frac{6032\zeta_3}{27} + \frac{64\pi^4}{45}
\nonumber
\\
&\hspace{-3em} - \frac{19408\pi^2}{243} - \frac{80324}{729} \bigg) L_S \Bigg]
+ C_FC_A^2 \Bigg[ - \frac{1936}{27} L_S^4 - \bigg( \frac{352\pi^2}{27} - \frac{28480}{81} \bigg) L_S^3 + \bigg( \frac{104\pi^2}{27}
\nonumber
\\
&\hspace{-3em} - \frac{88\pi^4}{45} - \frac{62012}{81} + 352 \zeta_3  \bigg) L_S^2 + \bigg( \frac{50344\pi^2}{243} - \frac{88\pi^4}{9} + \frac{556042}{729} + 384\zeta_5 + \frac{176\pi^2\zeta_3}{9}
\nonumber
\\
&\hspace{-3em} - \frac{36272\zeta_3}{27} \bigg) L_S \Bigg]
+ C_F^2n_f \Bigg[ \frac{256}{9} L_S^5 - \frac{640}{9} L_S^4 + \bigg( \frac{32\pi^2}{3} + \frac{1504}{27} \bigg) L_S^3 + \bigg( \frac{5620}{81} - \frac{856\pi^2}{27}
\nonumber
\\
&\hspace{-3em} - \frac{160\zeta_3}{9} \bigg) L_S^2 + \bigg( \frac{608\zeta_3}{9} + \frac{56\pi^4}{45} + \frac{152\pi^2}{27} - \frac{3422}{27} \bigg) L_S \Bigg]+ C_F^2C_A \Bigg[ - \frac{1408}{9} L_S^5
\nonumber
\\
&\hspace{-3em} + \Bigg( \frac{4288}{9} - \frac{64\pi^2}{3} \Bigg) L_S^4 + \bigg( 448 \zeta_3 - \frac{176\pi^2}{3} - \frac{12928}{27} \bigg) L_S^3 + \bigg( \frac{5092\pi^2}{27} - \frac{2288\zeta_3}{9} - \frac{152\pi^4}{15}
\nonumber
\\
&\hspace{-3em} + \frac{17120}{81} \bigg) L_S^2
+ \bigg( 56 \pi ^2 \zeta_3 - \frac{44\pi^4}{9} - \frac{1616\pi^2}{27} \bigg) L_S \Bigg] + C_F^3 \Bigg[ -\frac{256}{3} L_S^6 - 32 \pi^2 L_S^4 - 4 \pi^4 L_S^2 \Bigg]
\nonumber
\\
&\hspace{-3em} + c_{3q}^S \, .
\label{eq:threeloopsoftfunction}
\end{align}
The three-loop scale-independent term $c_{3q}^S$ is not precisely known at the moment. Its calculation is under active investigation \cite{Chen:2020dpk, Baranowski:2021gxe, Baranowski:2022khd}. In Ref.~\cite{Bruser:2018rad}, this term was extracted through a numeric fit to the fixed-order thrust distribution. The value (with large uncertainties) reads
\begin{equation}
  c^S_{3q}=-19988 \pm 1440 (\text{stat.}) \pm 4000 (\text{syst.}) \, .
\end{equation}

The results for the gluon soft function can be obtained from the quark one employing the non-Abelian exponential theorem \cite{Sterman:1981jc, Gatheral:1983cz, Frenkel:1984pz}. Up to three loops, they are related by the Casimir scaling
\begin{equation}
   \ln \left[ \tilde{s}^{g}(L_S,\mu) \right] = \frac{C_A}{C_F} \, \ln \left[ \tilde{s}^{q}(L_S,\mu) \right] .
\end{equation}
Hence the expansion coefficients for the gluon soft function are
\begin{align}
\tilde{s}^{g(1)}(L_S) &= C_A \left(-8 L_S^2-\pi ^2\right) ,
\nonumber
\\
\tilde{s}^{g(2)}(L_S) &= C_An_f \Bigg[ -\frac{32}{9} L_S^3+\frac{80}{9} L_S^2-\bigg(\frac{8 \pi^2}{9} +\frac{224}{27} \bigg)L_S+\frac{77 \pi^2}{27}+
\frac{40}{81} -\frac{52 \zeta _3}{9} \Bigg]
\nonumber
\\
&\hspace{-3em} + C_A^2 \Bigg[ 32 L_S^4+\frac{176}{9} L_S^3+\bigg(\frac{32 \pi ^2}{3}-\frac{536}{9}\bigg)L_S^2 + \bigg(\frac{44 \pi^2}{9}
+\frac{1616}{27} -56 \zeta _3 \bigg) L_S +\frac{286 \zeta _3}{9}
\nonumber
\\
&\hspace{-3em} +\frac{43 \pi ^4}{30}-\frac{871 \pi^2}{54}-\frac{2140}{81} \Bigg]
\, ,
\nonumber
\\
\tilde{s}^{g(3)}(L_S) &= C_An_f^2 \Bigg[ -\frac{64}{27} L_S^4 + \frac{640}{81} L_S^3 - \bigg (\frac{32\pi^2}{27} + \frac{800}{81} \bigg) L_S^2 + \bigg( \frac{64\pi^2}{9} - \frac{3200}{729} - \frac{64\zeta_3}{9} \bigg) L_S \Bigg]
\nonumber
\\
&\hspace{-3em} + C_F C_A n_f \Bigg[ -\frac{32}{3} L_S^3 +\left(\frac{220}{3}-64 \zeta_3\right) L_S^2 + \left(\frac{608 \zeta _3}{9}+\frac{16 \pi ^4}{45}-\frac{8 \pi^2}{3}-\frac{3422}{27}\right) L_S \Bigg]
\nonumber
\\
&\hspace{-3em}
+ C_A^2 n_f \Bigg[ \frac{256}{9} L_S^5 -\frac{1216}{27} L_S^4 +\left(\frac{352 \pi^2}{27}-\frac{3872}{81}\right) L_S^3 +\left(\frac{416 \zeta _3}{9}-\frac{664 \pi^2}{27}+\frac{16088}{81}\right) L_S^2
\nonumber
\\
&\hspace{-3em} +\left(\frac{6032 \zeta _3}{27}+\frac{104
   \pi ^4}{45}-\frac{17392 \pi ^2}{243}-\frac{80324}{729}\right) L_S \Bigg]
   + C_A^3 \Bigg[-\frac{256}{3} L_S^6 -\frac{1408}{9} L_S^5
\nonumber
\\
&\hspace{-3em}
  +\left(\frac{10928}{27}-\frac{160 \pi^2}{3}\right) L_S^4 +\left(448 \zeta _3-\frac{1936 \pi^2}{27}-\frac{10304}{81}\right) L_S^3 +\bigg(\frac{880 \zeta _3}{9}-\frac{724 \pi^4}{45}
\nonumber
\\
&\hspace{-3em}  +\frac{1732 \pi ^2}{9}-\frac{4988}{9} \bigg) L_S^2 +\bigg(\frac{680 \pi^2\zeta _3}{9}-\frac{36272 \zeta _3}{27}+384 \zeta _5-\frac{44 \pi^4}{3}+\frac{35800 \pi ^2}{243}
\nonumber
\\
&\hspace{-3em}
+\frac{556042}{729}\bigg) L_S \Bigg]+ c_{3g}^S \, ,
\end{align}
where
\begin{align}\label{exp:csg}
   c^S_{3g} &= C_A C_F n_f \left(-\frac{52}{9} \pi ^2 \zeta _3+\frac{77 \pi^4}{27}+\frac{40 \pi^2}{81}\right) + C_A^2 n_f \left(\frac{52 \pi ^2 \zeta_3}{9}-\frac{77 \pi^4}{27}-\frac{40 \pi ^2}{81}\right)
\nonumber
\\
&\hspace{1em}
 + C_A^3 \left(-\frac{286}{9} \pi ^2 \zeta_3-\frac{11 \pi ^6}{10}
 +\frac{871 \pi ^4}{54}+\frac{2140 \pi ^2}{81}\right)
 + C_A^2 C_F \bigg(\frac{286 \pi ^2 \zeta _3}{9}+\frac{14 \pi ^6}{15}
\nonumber
\\
&\hspace{1em}
 -\frac{871 \pi^4}{54} -\frac{2140 \pi ^2}{81}\bigg) + C_AC_F^2 \frac{1}{6} \pi ^6 + \frac{C_A}{C_F}c^S_{3q} \, .
\end{align}

\section{Anomalous dimensions}
\label{sec:app:Anomalous dimensions}

In this Appendix we list the expressions of the various anomalous dimensions appearing in the resummation formula.

For the cusp anomalous dimensions, we write
\begin{equation}
\Gamma_{\text{cusp}}^{q}(\alpha_s) = C_F \left[ \gamma_{\text{cusp}}(\alpha_s) + \delta\gamma_{\text{cusp}}^q(\alpha_s) \right] , \quad \Gamma_{\text{cusp}}^{g}(\alpha_s) = C_A \left[ \gamma_{\text{cusp}}(\alpha_s) + \delta\gamma_{\text{cusp}}^g(\alpha_s) \right] .
\end{equation}
Up to three loops the cusp anomalous dimensions satisfy Casimir scaling, so that $\delta\gamma_{\text{cusp}}^q(\alpha_s)$ and $\delta\gamma_{\text{cusp}}^g(\alpha_s)$ only start at $\alpha_s^4$. We define the expansion as
\begin{equation}
\gamma_{\text{cusp}}(\alpha_s) = \sum_{n=0} \gamma^{(n)}_{\text{cusp}}  \left( \frac{\alpha_s}{4\pi} \right)^{n+1} \, , \quad \delta\gamma_{\text{cusp}}^{q,g}(\alpha_s) = \sum_{n=3} \delta\gamma^{q,g(n)}_{\text{cusp}}  \left( \frac{\alpha_s}{4\pi} \right)^{n+1}
\end{equation}
The expansion coefficients are \cite{Moch:2004pa, Henn:2019swt, vonManteuffel:2020vjv, Herzog:2018kwj}
\begin{align}
\gamma_{\text{cusp}}^{(0)} &= 4 \, , \nonumber
\\
\gamma_{\text{cusp}}^{(1)} &= -\frac{1}{3} 4 \pi ^2 C_A+\frac{268 C_A}{9}-\frac{80 n_f T_F}{9} \, , \nonumber
\\
\gamma_{\text{cusp}}^{(2)} &= -\frac{16 n_f^2}{27}-\frac{208 n_f \zeta (3)}{3}+\frac{80 \pi ^2 n_f}{9}-\frac{1276 n_f}{9}+264 \zeta_3+\frac{44 \pi ^4}{5}-\frac{536 \pi ^2}{3} + 1470n_f \, , \nonumber
\\
\gamma_{\text{cusp}}^{(3)} &=
n_f^3 \left(\frac{64 \zeta_3}{27}-\frac{32}{81}\right)+n_f^2 \left(\frac{4160 \zeta_3}{27}-\frac{304 \pi ^2}{81}-\frac{104 \pi ^4}{135}+\frac{17875}{243}\right)
+n_f \left(\frac{416 \pi ^2 \zeta_3}{3}\right.\nonumber\\
&\left.-\frac{153920 \zeta_3}{27}+\frac{19504 \zeta_5}{9}+\frac{12800 \pi ^2}{27}-\frac{616 \pi ^4}{45}-\frac{344345}{81}\right)-432 \zeta_3^2-528 \pi ^2 \zeta_3\nonumber\\
&+20944 \zeta_3-10824 \zeta_5+\frac{2706 \pi ^4}{5}+\frac{84278}{3}-\frac{44200 \pi ^2}{9}-\frac{2504 \pi ^6}{105} \, , \nonumber
\\
\delta\gamma_{\text{cusp}}^{q(3)} &= n_f \left(-\frac{80 \zeta_3}{9}-\frac{400 \zeta_5}{9}+\frac{40 \pi ^2}{9}\right)-720 \zeta_3^2+80 \zeta_3+2200 \zeta_5-40 \pi ^2-\frac{124 \pi ^6}{63} \, , \nonumber
\\
\delta\gamma_{\text{cusp}}^{g(3)} &= n_f \left(-\frac{80 \zeta_3}{3}-\frac{400 \zeta_5}{3}+\frac{40 \pi ^2}{3}\right)-2160 \zeta_3^2+240 \zeta_3+6600 \zeta_5-120 \pi ^2 -\frac{124 \pi ^6}{21} \, .
\end{align}
As far as we know, there is no complete results for the five-loop cusp anomalous dimensions. In this paper, we make use of the approximate results estimated in \cite{Herzog:2018kwj},
\begin{align}
\gamma_{\text{cusp}}^{(4)} + \delta\gamma_{\text{cusp}}^{q(4)} &= 50000 \pm 40000 \,, \nonumber
\\
\gamma_{\text{cusp}}^{(4)} + \delta\gamma_{\text{cusp}}^{g(4)} &= 30000 \pm 60000 \,.
\end{align}

The anomalous dimension for the quark Yukawa coupling reads
\begin{equation}
\gamma_y = - \sum_{n=0} \left( \frac{\alpha_s}{4 \pi} \right)^{n+1} \gamma^{(n)}_{y} \, ,
\end{equation}
where \cite{Vermaseren:1997fq, Chetyrkin:1997dh}
\begin{align}
\gamma_y^{(0)} &= 6 C_F \, , \nonumber
\\
\gamma_y^{(1)} &= \frac{97 C_A C_F}{3}-\frac{10 C_F n_f}{3}+3 C_F^2 \, , \nonumber
\\
\gamma_y^{(2)} &= -48 \zeta_3 C_A C_F n_f-\frac{556}{27} C_A C_F n_f-\frac{129}{2} C_A C_F^2+\frac{11413}{54} C_A^2 C_F+48 \zeta_3 C_F^2 n_f \nonumber
\\
&- 46 C_F^2 n_f-\frac{70}{27} C_F n_f^2+129 C_F^3 \, , \nonumber
\\
\gamma_y^{(3)} &=
n_f^3 \left(\frac{128 \zeta_3}{27}-\frac{664}{243}\right)
+n_f^2 \left(\frac{1600 \zeta_3}{9}-\frac{32 \pi ^4}{27}+\frac{10484}{243}\right)
+n_f \left(-\frac{68384 \zeta_3}{9}\right. \nonumber
\\
&\left.+\frac{36800 \zeta_5}{9}+\frac{176 \pi ^4}{9}-\frac{183446}{27}\right)+\frac{271360 \zeta_3}{27}-17600 \zeta_5+\frac{4603055}{81} \, .
\end{align}

The anomalous dimension for the Wilson coefficient $C_t$ is ~
\begin{equation}
\gamma_{t}(\alpha_s)=\sum_{n=0} \left( \dfrac{\alpha_s}{4 \pi} \right)^{n+1} \gamma^{(n)}_{t}  \, ,
\end{equation}
where \cite{Spiridonov:1984br,Kramer:1996iq,Chetyrkin:1997un,Schroder:2005hy,Chetyrkin:2005ia}
\begin{align}
\gamma_t^{(0)} &=0 \, , \nonumber
\\
\gamma_t^{(1)} &= \frac{40}{3} C_A n_f T_F-\frac{1}{3} 68 C_A^2+8 C_F n_f T_F \, , \nonumber
\\
\gamma_t^{(2)} &= -\frac{1}{27} 650 n_f^2+\frac{10066 n_f}{9}-5714 \, , \nonumber
\\
\gamma_t^{(2)} &= -\frac{2186 n_f^3}{243}+n_f^2 \left(-\frac{12944 \zeta_3}{27}-\frac{50065}{27}\right)+n_f \left(\frac{13016 \zeta_3}{9}+\frac{1078361}{27}\right) \nonumber
\\
&\hspace{8cm} -21384 \zeta_3-149753 \, .
\end{align}

The anomalous dimension for the jet functions are
\begin{equation}
\gamma_j^{q,g}(\alpha_s) = \sum_{n=0} \left( \frac{\alpha_s}{4 \pi} \right)^{n+1} \gamma^{q,g(n)}_j \, ,
\end{equation}
where \cite{Becher:2006mr, Duhr:2022cob}
\begin{equation}
\begin{split}
\gamma_j^{q(0)} &= -3 C_F\,,\\
\gamma_j^{q(1)} &= \frac{8 \pi ^2 n_f}{27}+\frac{484 n_f}{81}+\frac{352 \zeta_3}{3}-\frac{4 \pi ^2}{3}-\frac{3610}{27} \,,\\
\gamma_j^{q(2)} &= -\frac{256}{81} \zeta_3 n_f^2-\frac{14272 \zeta_3 n_f}{81}-\frac{80}{243} \pi ^2 n_f^2+\frac{13828 n_f^2}{2187}+\frac{1172 \pi ^4
   n_f}{1215}+\frac{1592 \pi ^2 n_f}{243}\\
   &+\frac{100984 n_f}{729}-\frac{25696 \zeta_5}{9}-\frac{9632 \pi ^2 \zeta_3}{81}+\frac{153136 \zeta
   (3)}{27}-\frac{6818 \pi ^4}{405}+\frac{7588 \pi ^2}{81}\\
   &-\frac{470183}{243} \,,\\
   \gamma_j^{q(3)} &= 4483.56\,,
\end{split}
\end{equation}
and \cite{Becher:2009th, Duhr:2022cob}
\begin{equation}
\begin{split}
\gamma_j^{g(0)}&=-\beta_0 \,,\\
\gamma_j^{g(1)}&= -\frac{2}{9} \pi ^2  n_f  C_A+\frac{184 n_f C_A}{27}+16 \zeta_3 C_A^2+\frac{11}{9} \pi ^2 C_A^2-\frac{1096 C_A^2}{27}+2 n_f C_F \,,\\
\gamma_j^{g(2)}&=-\frac{304}{9} n_f \zeta_3 C_A C_F-\frac{8}{45} \pi ^4 n_f C_A C_F-\frac{2}{3} \pi ^2 n_f C_A C_F+\frac{4145}{54} n_f C_A C_F-\frac{112}{27} n_f^2 \zeta_3 C_A\\
&+\frac{1811 n_f^2 C_A}{1458}+\frac{20}{81} \pi ^2 n_f^2 C_A-\frac{8}{27} n_f \zeta_3 C_A^2+\frac{77}{135} \pi ^4 n_f C_A^2-\frac{1306}{243} \pi ^2 n_f C_A^2\\
&+\frac{42557 n_f C_A^2}{1458}-\frac{64}{9} \pi ^2 \zeta_3 C_A^3+260 \zeta_3 C_A^3-112 \zeta_5 C_A^3+\frac{6217}{243} \pi ^2 C_A^3-\frac{583}{270} \pi ^4 C_A^3\\
&-\frac{331153 C_A^3}{1458}-\frac{11 n_f^2 C_F}{9}-n_f C_F^2\,,\\
\gamma_j^{g(3)}&=26138.7\,.
\end{split}
\end{equation}

The hard anomalous dimensions are
\begin{equation}
\gamma_H^{q,g}(\alpha_s)=\sum_{n=0} \left( \frac{\alpha_s}{4 \pi} \right)^{n+1}  \gamma^{q,g(n)}_H \, ,
\end{equation}
where the coefficients up to three loops can be obtained from Eq.~\eqref{eq:hardfunctionExplicitHgg} and Eq.~\eqref{eq:hardfunctionExplicitHqq}. For the gluon case, they read \cite{Dawson:1990zj, Djouadi:1991tka, Harlander:2000mg, Moch:2005tm, Gehrmann:2005pd, Gehrmann:2010ue, Lee:2022nhh}
\begin{equation}
\begin{split}
\gamma_H^{g(0)}&=0\,,\\
\gamma_H^{g(1)}&=-\frac{ 2 \pi ^2  n_f}{3}  -\frac{152  {n_f}}{9}+36 \zeta_3+11 \pi ^2-\frac{160}{3}\,,\\
\gamma_H^{g(2)}&=-\frac{112 n_f^2 \zeta_3}{9}+\frac{20 \pi ^2 n_f^2}{27}+\frac{7714 n_f^2}{243}+\frac{920 n_f \zeta_3 }{9}+\frac{214 \pi ^4 n_f}{45}-\frac{1270 \pi ^2 n_f}{27}\\
&-\frac{76256 n_f}{81}-120 \pi ^2 \zeta_3+2196 \zeta_3-864 \zeta_5+\frac{6109 \pi ^2}{9}-\frac{319 \pi ^4}{5}+\frac{37045}{27}\,,\\
\gamma_H^{g(3)}&=
n_f^3 \left(\frac{400 \zeta_3}{81}+\frac{8 \pi ^2}{81}-\frac{64 \pi ^4}{405}+\frac{38426}{2187}\right)
+n_f^2 \left(\frac{368 \pi ^2 \zeta_3}{9}-\frac{49342 \zeta_3}{81}\right.\\
&\left.+\frac{8000 \zeta_5}{9}+\frac{19675 \pi ^2}{486}-\frac{2474 \pi ^4}{405}+\frac{3605645}{1944}\right)
+n_f \left(\frac{32 \zeta_3^2}{3}-504 \pi ^2 \zeta_3\right.\\
&\left.+\frac{528362 \zeta_3}{27}-\frac{90140 \zeta_5}{9}+\frac{52153 \pi ^4}{135}-\frac{262193 \pi ^2}{162}-\frac{7927313}{216}-\frac{54917 \pi ^6}{2835}\right)\\
&+21384 \zeta_3^2+\frac{2004 \pi ^4 \zeta_3}{5}-576 \pi ^2 \zeta_3+\frac{119536 \zeta_3}{3}+1152 \pi ^2 \zeta_5\\
&-62796 \zeta_5-22734 \zeta_7+\frac{18051 \pi ^6}{70}+\frac{1041691 \pi ^2}{54}-\frac{123029 \pi ^4}{30}\\
&+\frac{5481844}{81} \, .
\end{split}
\end{equation}
For the quark case, they are given as \cite{Dawson:1990zj, Harlander:2003ai, Gehrmann:2014vha, Chakraborty:2022yan}
\begin{equation}
\begin{split}
\gamma_H^{q(0)}&=0 \,,\\
\gamma_H^{q(1)}&=\frac{8 \pi ^2 n_f}{9}+\frac{160 n_f }{81}+\frac{368 \zeta_3}{3}-\frac{68 \pi ^2}{9}-\frac{352}{27} \,,\\
\gamma_H^{q(2)}&=-\frac{64 n_f^2 \zeta_3 }{81}-\frac{ 80 \pi ^2 n_f^2}{81} +\frac{11776 n_f^2}{2187}-\frac{23584 n_f \zeta_3 }{81}+\frac{136 \pi ^4 n_f}{1215}+\frac{9128 \pi ^2 n_f}{243}\\
&-\frac{47192 n_f}{729}+\frac{164144 \zeta_3}{27}-\frac{30656 \zeta_5 }{9}-\frac{9760 \pi ^2 \zeta_3 }{81}-\frac{10124 \pi ^2}{81}+\frac{213772}{243}\\
&-\frac{4132 \pi ^4}{405} \,,\\
\gamma_s^{q(3)}&=
n_f^3 \left(-\frac{2240 \zeta_3}{729}-\frac{32 \pi ^2}{243}-\frac{128 \pi ^4}{3645}+\frac{95744}{19683}\right)
+n_f^2 \left(-\frac{1}{243} 2816 \pi ^2 \zeta_3\right.\\
&\left.+\frac{18848 \zeta_3}{729}+\frac{25984 \zeta_5}{81}+\frac{6856 \pi ^4}{10935}-\frac{130486 \pi ^2}{2187}+\frac{126461}{4374}\right)+n_f \left(-\frac{110848 \zeta_3^2}{27}\right.\\
&\left.+\frac{115504 \pi ^2 \zeta_3}{243}-\frac{3066064 \zeta_3}{243}+\frac{59488 \zeta_5}{9}+\frac{29264 \pi ^4}{729}+\frac{755786 \pi ^2}{729}-\frac{1641457}{486}\right.\\
&\left.-\frac{975668 \pi ^6}{229635}\right)+\frac{175712 \zeta_3^2}{9}+\frac{992696 \pi ^4 \zeta_3}{3645}+\frac{1365152 \zeta_3}{9}+\frac{243872 \pi ^2 \zeta_5}{81}\\
&+\frac{2888332 \zeta_7}{27}-\frac{81584 \pi ^2 \zeta_3}{9}-\frac{2158832 \zeta_5}{9}+\frac{2058794 \pi ^6}{25515}-\frac{362842 \pi ^2}{243}\\
&+\frac{19161124}{729}-\frac{1095218 \pi ^4}{1215}\,.
\end{split}
\end{equation}

Due to the RG invariance of physical observables, the soft anomalous dimensions satisfy the consistency relations
\begin{equation}
\begin{split}
\gamma_s^q&=\gamma_H^q +\gamma_y -2\gamma_j^q  \,, \\
\gamma_s^g&=\gamma_H^g +\gamma_t+\frac{\beta}{\alpha_s}  -2\gamma_j^g \,.
\end{split}
\end{equation}
We expand them in $\alpha_s$
\begin{equation}
\begin{split}
\gamma_s^{q,g}(\alpha_s)=\sum_{n=0} \left(   \frac{\alpha_s}{4 \pi} \right)^{n+1} \gamma^{q,g(n)}_s \,,
\end{split}
\end{equation}
where
\begin{equation}
\begin{split}
\gamma_s^{q(0)}&=0 \,,\\
\gamma_s^{q(1)}&=  \frac{8 \pi ^2 n_f}{27}-\frac{448 n_f}{81}-112 \zeta_3-\frac{44 \pi ^2}{9}+\frac{3232}{27}\,,\\
\gamma_s^{q(2)}&= \frac{448 \zeta_3 n_f^2}{81}+\frac{13600 \zeta_3 n_f}{81}-\frac{80}{243} \pi ^2 n_f^2-\frac{8320 n_f^2}{2187}-\frac{736 \pi ^4 n_f}{405}+\frac{5944
   \pi ^2 n_f}{243}\\
   &-\frac{129496 n_f}{729}+2304 \zeta_5+\frac{352 \pi ^2 \zeta_3}{3}-5264 \zeta_3+\frac{352 \pi ^4}{15}-\frac{25300 \pi
   ^2}{81}+\frac{547124}{243}\,,\\
 \gamma_s^{q(3)}&= -5350.4\,,
\end{split}
\end{equation}
and
\begin{equation}
\begin{split}
\gamma_s^{g(0)}&=0 \,,\\
\gamma_s^{g(1)}&= \frac{2 \pi ^2 n_f}{3}-\frac{112 n_f}{9}-252 \zeta_3-11 \pi ^2+\frac{808}{3}\,,\\
\gamma_s^{g(2)}&= \frac{112 n_f^2 \zeta_3}{9}-\frac{20 \pi ^2 n_f^2}{27}-\frac{2080 n_f^2}{243}+\frac{3400 n_f \zeta_3}{9}+\frac{1486 \pi ^2 n_f}{27}-\frac{184 \pi ^4 n_f}{45}\\
&-\frac{32374 n_f}{81}+264 \pi ^2 \zeta_3-11844 \zeta_3+5184 \zeta_5+\frac{264 \pi ^4}{5}-\frac{6325 \pi ^2}{9}+\frac{136781}{27}\,,\\
\gamma_s^{g(3)}&=-14715.4\,.
\end{split}
\end{equation}
Up to three loops, the quark and gluon soft anomalous dimensions satisfy the Casimir scaling $ \gamma_s^g/C_A=\gamma_s^q/C_F$. However, starting at four loops, due to the emergence of new Casimir operators, the relation must be generalized as appropriate \cite{Moch:2018wjh, Duhr:2022cob}.

Finally, the beta function coefficients are \cite{vanRitbergen:1997va, Czakon:2004bu}
\begin{equation}
\begin{split}
\beta_{0}&=\frac{11 C_A}{3}-\frac{4 n_f T_F}{3}\,,\\
\beta_{1}&=\frac{34 C_A^2}{3}-\frac{20 C_A n_f T_F}{3}-4 C_F n_f T_F\,,\\
\beta_{2}&=\frac{325 n_f^2}{54}-\frac{5033 n_f}{18}+\frac{2857}{2}\,,\\
\beta_{3}&=\frac{1093 n_f^3}{729}+n_f^2 \left(\frac{6472 \zeta_3}{81}+\frac{50065}{162}\right)+n_f \left(-\frac{6508 \zeta_3}{27}-\frac{1078361}{162}\right)+3564 \zeta_3\\
&+\frac{149753}{6}\,,\\
\beta_{4}&=n_f^4 \left(\frac{1205}{2916}-\frac{152 \zeta_3}{81}\right)
+n_f^3 \left(-\frac{48722 \zeta_3}{243}+\frac{460 \zeta_5}{9}+\frac{809 \pi ^4}{1215}-\frac{630559}{5832}\right)\\
&+n_f^2 \left(\frac{698531 \zeta_3}{81}-\frac{381760 \zeta_5}{81}-\frac{5263 \pi ^4}{405}+\frac{25960913}{1944}\right)
+n_f \left(-\frac{4811164 \zeta_3}{81}\right.\\
&\left.+\frac{1358995 \zeta_5}{27}+\frac{6787 \pi ^4}{108}-\frac{336460813}{1944}\right)+\frac{621885 \zeta_3}{2}-288090 \zeta_5+\frac{8157455}{16}\\
&-\frac{9801 \pi ^4}{20}\,.
\end{split}
\end{equation}

\clearpage

\bibliographystyle{JHEP}
\bibliography{thrust.bib}

\end{document}